\documentclass[
    aps,
    prd,
    reprint,
    amsmath,
    amssymb,
    nofootinbib,
    superscriptaddress,
    floatfix
]{revtex4-2}

\usepackage{graphicx}
\usepackage{float}
\usepackage{siunitx}
\usepackage{tabularx}
\usepackage[utf8x]{inputenc}
\usepackage{gensymb}

\usepackage[ngerman, english]{babel} % load characters 

\usepackage{xcolor}

\definecolor{tab-orange}{RGB}{255, 128, 14}
\definecolor{tab-blue}{RGB}{0, 107, 164}

\usepackage[
    colorlinks=true,
    allcolors=tab-blue
]{hyperref}

\usepackage[all]{hypcap}

\usepackage[
    noabbrev,
    capitalise,
    nameinlink,
]{cleveref}

\newcommand{\gev}{\giga\electronvolt}
\newcommand{\mev}{\mega\electronvolt}

\def\ue4{$|U_{e4}|^2$}
\def\um4{$|U_{\mu4}|^2$}
\def\ut4{$|U_{\tau4}|^2$}

\def\nomathue4{|U_{e4}|^2}
\def\nomathum4{|U_{\mu4}|^2}
\def\nomathut4{|U_{\tau4}|^2}

\ifdefined\rm
\renewcommand{\rm}{\mathrm}
\else
\newcommand{\rm}{\mathrm}
\fi

\begin{document}

\title{Search for Heavy Neutral Leptons with IceCube DeepCore}


%%%%%%%%%%%%%%%%%%%% Authorlist %%%%%%%%%%%%%%%%%%%%

\affiliation{III. Physikalisches Institut, RWTH Aachen University, D-52056 Aachen, Germany}
\affiliation{Department of Physics, University of Adelaide, Adelaide, 5005, Australia}
\affiliation{Dept. of Physics and Astronomy, University of Alaska Anchorage, 3211 Providence Dr., Anchorage, AK 99508, USA}
\affiliation{Dept. of Physics, University of Texas at Arlington, 502 Yates St., Science Hall Rm 108, Box 19059, Arlington, TX 76019, USA}
\affiliation{School of Physics and Center for Relativistic Astrophysics, Georgia Institute of Technology, Atlanta, GA 30332, USA}
\affiliation{Dept. of Physics, Southern University, Baton Rouge, LA 70813, USA}
\affiliation{Dept. of Physics, University of California, Berkeley, CA 94720, USA}
\affiliation{Lawrence Berkeley National Laboratory, Berkeley, CA 94720, USA}
\affiliation{Institut f{\"u}r Physik, Humboldt-Universit{\"a}t zu Berlin, D-12489 Berlin, Germany}
\affiliation{Fakult{\"a}t f{\"u}r Physik {\&} Astronomie, Ruhr-Universit{\"a}t Bochum, D-44780 Bochum, Germany}
\affiliation{Universit{\'e} Libre de Bruxelles, Science Faculty CP230, B-1050 Brussels, Belgium}
\affiliation{Vrije Universiteit Brussel (VUB), Dienst ELEM, B-1050 Brussels, Belgium}
\affiliation{Dept. of Physics, Simon Fraser University, Burnaby, BC V5A 1S6, Canada}
\affiliation{Department of Physics and Laboratory for Particle Physics and Cosmology, Harvard University, Cambridge, MA 02138, USA}
\affiliation{Dept. of Physics, Massachusetts Institute of Technology, Cambridge, MA 02139, USA}
\affiliation{Dept. of Physics and The International Center for Hadron Astrophysics, Chiba University, Chiba 263-8522, Japan}
\affiliation{Department of Physics, Loyola University Chicago, Chicago, IL 60660, USA}
\affiliation{Dept. of Physics and Astronomy, University of Canterbury, Private Bag 4800, Christchurch, New Zealand}
\affiliation{Dept. of Physics, University of Maryland, College Park, MD 20742, USA}
\affiliation{Dept. of Astronomy, Ohio State University, Columbus, OH 43210, USA}
\affiliation{Dept. of Physics and Center for Cosmology and Astro-Particle Physics, Ohio State University, Columbus, OH 43210, USA}
\affiliation{Niels Bohr Institute, University of Copenhagen, DK-2100 Copenhagen, Denmark}
\affiliation{Dept. of Physics, TU Dortmund University, D-44221 Dortmund, Germany}
\affiliation{Dept. of Physics and Astronomy, Michigan State University, East Lansing, MI 48824, USA}
\affiliation{Dept. of Physics, University of Alberta, Edmonton, Alberta, T6G 2E1, Canada}
\affiliation{Erlangen Centre for Astroparticle Physics, Friedrich-Alexander-Universit{\"a}t Erlangen-N{\"u}rnberg, D-91058 Erlangen, Germany}
\affiliation{Physik-department, Technische Universit{\"a}t M{\"u}nchen, D-85748 Garching, Germany}
\affiliation{D{\'e}partement de physique nucl{\'e}aire et corpusculaire, Universit{\'e} de Gen{\`e}ve, CH-1211 Gen{\`e}ve, Switzerland}
\affiliation{Dept. of Physics and Astronomy, University of Gent, B-9000 Gent, Belgium}
\affiliation{Dept. of Physics and Astronomy, University of California, Irvine, CA 92697, USA}
\affiliation{Karlsruhe Institute of Technology, Institute for Astroparticle Physics, D-76021 Karlsruhe, Germany}
\affiliation{Karlsruhe Institute of Technology, Institute of Experimental Particle Physics, D-76021 Karlsruhe, Germany}
\affiliation{Dept. of Physics, Engineering Physics, and Astronomy, Queen's University, Kingston, ON K7L 3N6, Canada}
\affiliation{Department of Physics {\&} Astronomy, University of Nevada, Las Vegas, NV 89154, USA}
\affiliation{Nevada Center for Astrophysics, University of Nevada, Las Vegas, NV 89154, USA}
\affiliation{Dept. of Physics and Astronomy, University of Kansas, Lawrence, KS 66045, USA}
\affiliation{Centre for Cosmology, Particle Physics and Phenomenology - CP3, Universit{\'e} catholique de Louvain, Louvain-la-Neuve, Belgium}
\affiliation{Department of Physics, Mercer University, Macon, GA 31207-0001, USA}
\affiliation{Dept. of Astronomy, University of Wisconsin{\textemdash}Madison, Madison, WI 53706, USA}
\affiliation{Dept. of Physics and Wisconsin IceCube Particle Astrophysics Center, University of Wisconsin{\textemdash}Madison, Madison, WI 53706, USA}
\affiliation{Institute of Physics, University of Mainz, Staudinger Weg 7, D-55099 Mainz, Germany}
\affiliation{Department of Physics, Marquette University, Milwaukee, WI 53201, USA}
\affiliation{Institut f{\"u}r Kernphysik, Universit{\"a}t M{\"u}nster, D-48149 M{\"u}nster, Germany}
\affiliation{Bartol Research Institute and Dept. of Physics and Astronomy, University of Delaware, Newark, DE 19716, USA}
\affiliation{Dept. of Physics, Yale University, New Haven, CT 06520, USA}
\affiliation{Columbia Astrophysics and Nevis Laboratories, Columbia University, New York, NY 10027, USA}
\affiliation{Dept. of Physics, University of Oxford, Parks Road, Oxford OX1 3PU, United Kingdom}
\affiliation{Dipartimento di Fisica e Astronomia Galileo Galilei, Universit{\`a} Degli Studi di Padova, I-35122 Padova PD, Italy}
\affiliation{Dept. of Physics, Drexel University, 3141 Chestnut Street, Philadelphia, PA 19104, USA}
\affiliation{Physics Department, South Dakota School of Mines and Technology, Rapid City, SD 57701, USA}
\affiliation{Dept. of Physics, University of Wisconsin, River Falls, WI 54022, USA}
\affiliation{Dept. of Physics and Astronomy, University of Rochester, Rochester, NY 14627, USA}
\affiliation{Department of Physics and Astronomy, University of Utah, Salt Lake City, UT 84112, USA}
\affiliation{Dept. of Physics, Chung-Ang University, Seoul 06974, Republic of Korea}
\affiliation{Oskar Klein Centre and Dept. of Physics, Stockholm University, SE-10691 Stockholm, Sweden}
\affiliation{Dept. of Physics and Astronomy, Stony Brook University, Stony Brook, NY 11794-3800, USA}
\affiliation{Dept. of Physics, Sungkyunkwan University, Suwon 16419, Republic of Korea}
\affiliation{Institute of Basic Science, Sungkyunkwan University, Suwon 16419, Republic of Korea}
\affiliation{Institute of Physics, Academia Sinica, Taipei, 11529, Taiwan}
\affiliation{Dept. of Physics and Astronomy, University of Alabama, Tuscaloosa, AL 35487, USA}
\affiliation{Dept. of Astronomy and Astrophysics, Pennsylvania State University, University Park, PA 16802, USA}
\affiliation{Dept. of Physics, Pennsylvania State University, University Park, PA 16802, USA}
\affiliation{Dept. of Physics and Astronomy, Uppsala University, Box 516, SE-75120 Uppsala, Sweden}
\affiliation{Dept. of Physics, University of Wuppertal, D-42119 Wuppertal, Germany}
\affiliation{Deutsches Elektronen-Synchrotron DESY, Platanenallee 6, D-15738 Zeuthen, Germany}

\author{R. Abbasi}
\affiliation{Department of Physics, Loyola University Chicago, Chicago, IL 60660, USA}
\author{M. Ackermann}
\affiliation{Deutsches Elektronen-Synchrotron DESY, Platanenallee 6, D-15738 Zeuthen, Germany}
\author{J. Adams}
\affiliation{Dept. of Physics and Astronomy, University of Canterbury, Private Bag 4800, Christchurch, New Zealand}
\author{S. K. Agarwalla}
\thanks{also at Institute of Physics, Sachivalaya Marg, Sainik School Post, Bhubaneswar 751005, India}
\affiliation{Dept. of Physics and Wisconsin IceCube Particle Astrophysics Center, University of Wisconsin{\textemdash}Madison, Madison, WI 53706, USA}
\author{J. A. Aguilar}
\affiliation{Universit{\'e} Libre de Bruxelles, Science Faculty CP230, B-1050 Brussels, Belgium}
\author{M. Ahlers}
\affiliation{Niels Bohr Institute, University of Copenhagen, DK-2100 Copenhagen, Denmark}
\author{J.M. Alameddine}
\affiliation{Dept. of Physics, TU Dortmund University, D-44221 Dortmund, Germany}
\author{N. M. Amin}
\affiliation{Bartol Research Institute and Dept. of Physics and Astronomy, University of Delaware, Newark, DE 19716, USA}
\author{K. Andeen}
\affiliation{Department of Physics, Marquette University, Milwaukee, WI 53201, USA}
\author{C. Arg{\"u}elles}
\affiliation{Department of Physics and Laboratory for Particle Physics and Cosmology, Harvard University, Cambridge, MA 02138, USA}
% ...

\date{\today}

\collaboration{IceCube Collaboration}
\noaffiliation

%%%%%%%%%%%%%%%%%%%% Abstract %%%%%%%%%%%%%%%%%%%%

\begin{abstract}

The observation of neutrino oscillations has established that neutrinos have non-zero masses. This phenomenon is not explained by the Standard Model of particle physics, but one viable explanation to this dilemma involves the existence of heavy neutral leptons in the form of right-handed neutrinos. This work presents the first search for heavy neutral leptons with the IceCube Neutrino Observatory. The standard three flavor neutrino model is extended by adding a fourth GeV-scale mass state allowing mixing with the $\tau$ sector through the parameter \ut4. The analysis is performed by searching for signatures of heavy neutral leptons that are directly produced via up-scattering of atmospheric $\nu_\tau$'s inside the IceCube detection volume.
Three heavy neutral lepton mass values, $m_4$, of \SI{0.3}{\gev}, \SI{0.6}{\gev}, and \SI{1.0}{\gev} are tested using ten years of data, collected between 2011 and 2021. No significant signal of heavy neutral leptons is observed for any of the tested masses. The resulting constraints for the mixing parameter are $\nomathut4 < 0.19\;(m_4 = \SI{0.3}{\gev})$, $\nomathut4 < 0.36\;(m_4 = \SI{0.6}{\gev})$, and $\nomathut4 < 0.40\;(m_4 = \SI{1.0}{\gev})$ at the \SI{90}{\percent} confidence level. This analysis serves as proof-of-concept for heavy neutral lepton searches in IceCube. The heavy neutral lepton event generator, developed in this work, and the analysis of the expected signatures lay the fundamental groundwork for future searches thereof.

\end{abstract}

\maketitle
\tableofcontents

%%%%%%%%%%%%%%%%%%%% Introduction %%%%%%%%%%%%%%%%%%%%

\section{\label{sec:introduction}Introduction}

The observation of neutrino flavor oscillations in solar~\cite{Cleveland:1998nv, Super-Kamiokande:2001ljr, Borexino:2008dzn, Gann:2021ndb}, atmospheric~\cite{Super-Kamiokande:1998kpq, IceCube:2013pav, ANTARES:2018rtf, IceCubeCollaboration:2023wtb, Arguelles:2022hrt}, and reactor experiments~\cite{DayaBay:2012fng, RENO:2012mkc, Vogel:2015wua} has yielded a plethora of evidence for non-zero neutrino masses~\cite{Esteban:2018azc}.
Despite decades of observations, the structure and origin of the neutrino mass term remains elusive~\cite{Mohapatra:2005wg}.
The Standard Model (SM) of particle physics, as originally formulated~\cite{Weinberg:1967tq, Weinberg:2004kv}, does not contain neutral right-handed singlet fermions, precluding a renormalizable neutrino-Higgs Yukawa interaction~\cite{Higgs:1964pj}.
This makes the existence of right-handed neutrinos, $N_R$, an appealing solution to the neutrino-mass puzzle.
In this scenario, in addition to participating in the Higgs mechanism for neutrinos through the coupling $y \bar L \tilde H N_R$, they admit an unconstrained Majorana mass term, $\frac{1}{2} m_R \bar N_R^c N_R$.
Here, $y$ is the Yukawa coupling, $L$ is the lepton doublet, $H$ is the Higgs doublet, and $N_R$ is the right-handed neutrino with Majorana mass $m_R$.
If neutrinos are Dirac particles, this latter term is exactly zero and lepton number is preserved, while if neutrinos are quasi-Dirac~\cite{Valle:1983dk}, as suggested by recent theories of quantum gravity~\cite{Ooguri:2016pdq, Arkani-Hamed:2007ryu, Gonzalo:2021zsp, Vafa:2024fpx}, this term is small and conducive to oscillations on astrophysical scales~\cite{Crocker:2001zs, Keranen:2003xd,Beacom:2003eu, Esmaili:2009fk, Esmaili:2012ac, Shoemaker:2015qul, Carloni:2022cqz}.
A third possibility is motivated by the see-saw mechanism~\cite{Minkowski:1977sc, Mohapatra:1979ia, GellMann:1980vs, Yanagida:1979as, Lazarides:1980nt, Mohapatra:1980yp, Schechter:1980gr, Cheng:1980qt, Foot:1988aq}, which aims to explain the smallness of neutrino masses, where, in its simplest realization, $m_R$ is GeV-scale or heavier.
This minimal see-saw scenario leaves us with stark prospects for experimental observation, where the mass of the heavy neutrino is typically too large to be directly produced or its couplings too small to produce detectable signals.
However, this is not the case in low-scale realizations of the see-saw mechanism~\cite{Mohapatra:1986bd, GonzalezGarcia:1988rw,Wyler:1982dd, Akhmedov:1995ip, Akhmedov:1995vm,Barry:2011wb, Zhang:2011vh}, where the approximate conservation of lepton masses guarantees the smallness of neutrino masses in a technically natural way~\cite{naturalness_tHooft, Vissani:1997ys}.
This scenario predicts the existence of quasi-Dirac heavy neutral leptons (HNLs) that have below-electroweak masses and mix with the known, light neutrino states.
If some of them are heavy (in the GeV regime) and some of them are light, as motivated in the $\nu$ Minimal Standard Model introduced in Refs.~\cite{ASAKA200517, ASAKA2005151},
% extending the SM by three right-handed neutrinos - where two of them are heavy and one of them is light - 
they could explain both the origin of neutrinos masses, while also providing a dark matter candidate. This work will therefore focus on HNLs in the GeV regime.

An HNL can be produced from the interactions of neutrinos through the weak force, since an HNL has a small component of left-handed neutrino flavors, leading to \textit{weaker-than-weak} interactions contributing to their production and decay.
Since HNL mixing with active neutrinos is presumed to be small, as it is yet unobserved, this leads to long HNL lifetimes barring the existence of new forces..
In this article, we consider the production of HNLs from high-energy atmospheric neutrinos interacting in the Antarctic ice, with subsequent decay of the HNLs, which could be detected with the IceCube Neutrino Observatory~\cite{IceCube:2016zyt}, a possibility first considered in Ref.~\cite{Coloma:2017ppo}.

The article is organized as follows.
First, we discuss potential HNL signals and proposals of how to search for them in \cref{sec:hnl_signals}.
In \cref{sec:detector}, we introduce the IceCube DeepCore detector, which is the primary instrument used in this search.
Following this, in \cref{sec:generator}, we describe the new event generator, used to simulate HNL events for this search.
Here also the detector response simulation and the HNL event re-weighting scheme are discussed.
In \cref{sec:filter_and_reco}, we describe the filtering steps leading to a neutrino-dominated analysis sample, including the reconstruction used in the analysis.
In particular, we explore the performance of reconstructing low-energy double-cascades with traditional likelihood methods and the resulting capability to identify them.
We find that the identification of these events against the background is challenging, leading a substantial difference between the model parameter space range where a single event is expected~\cite{Coloma:2017ppo, Boiarska:2021yho} to and our resulting limits.
In \cref{sec:analysis_methods,sec:results}, we describe the analysis methods and the results, respectively. 
% we describe the analysis methods and in \cref{sec:results} we present the results of this work.
Finally, in \cref{sec:conclusion} we conclude and discuss prospects for future HNL searches in IceCube.

%%%%%%%%%%%%%%%%%%%% Heavy Neutral Lepton Signals %%%%%%%%%%%%%%%%%%%%

\section{\label{sec:hnl_signals}Heavy Neutral Lepton Signals}

Searches for HNLs have been pursued since almost half a century~\cite{Abdullahi:2022jlv} based on various underlying models~\cite{Fernandez-Martinez:2023phj}. On top of the HNL production from protons interacting on a target or a beam dump, many searches utilized the fact that the produced HNL propagates undisturbed and unobserved before decaying back into SM particles, leading to spurious decay signatures or displaced vertices. Noteworthy experimental results at extracted beamlines are from \textit{CHARM}~\cite{CHARM:1983ayi, Orloff:2002de, Boiarska:2021yho}, \textit{NOMAD}~\cite{NOMAD:2001eyx}, and \textit{BEBC}~\cite{Barouki:2022bkt, BEBC_OG_COOPERSARKAR1985207}. At colliders, searches for HNLs are performed using prompt and displaced decays such as at with \textit{ATLAS}~\cite{ATLAS:2019kpx, atlas_2022_HNL_PhysRevLett.131.061803}, \textit{CMS}~\cite{CMS:2018iaf, CMS:2022fut}, and \textit{LHCb}~\cite{LHCb:2020wxx}, or using hadronic Z$^0$ decays like with \textit{DELPHI}~\cite{DELPHI:1996qcc}. For a thorough review of all searches for feebly-interacting particles, please refer to Ref.~\cite{Agrawal:2021dbo}.

Concerning searches for HNLs in atmospheric neutrino experiments, in addition to previously mentioned Ref.~\cite{Coloma:2017ppo}, Refs.~\cite{Boiarska:2021yho,Coloma:2019htx,Arguelles:2019ziu} also propose searches leveraging the Production of HNLs from the decay of mesons. In particular, Ref.~\cite{Boiarska:2021yho} proposes to look for HNLs using down-going events in IceCube and KM3NeT, an ocean-based neutrino telescope.
We do not follow these proposals because of the large downward-going muon background at these energies and the systematic uncertainties associated with D-meson production in the atmosphere.
In particular, when considering only upward-going events, those that traverse through the Earth, and HNLs in the GeV-scale, the production of meson decay is a sub-leading production mechanism compared to the one proposed in Ref.~\cite{Coloma:2017ppo}.
An additional unique advantage of looking for upward-going events is the access to a relatively large flux of $\nu_\tau$'s produced from the oscillation of $\nu_\mu$'s with energies in the tens of GeVs~\cite{IceCubeCollaboration:2023wtb}, where the neutrino cross-section is well-described by deep-inelastic scattering (DIS)~\cite{Formaggio:2012cpf}.
Given this unique advantage, and the fact that the mixing of $\nu_e$ and $\nu_\mu$ have been more thoroughly explored elsewhere~\cite{ParticleDataGroup:2024cfk}, we focus on HNLs whose largest active component is $\nu_\tau$ mixing.
Within our setup, Ref.~\cite{Coloma:2017ppo} predicted IceCube to detect one event in \SI{6}{years} of data taking, when the mixing was $\nomathut4\sim10^{-4}$ for a mass of \SI{1.0}{\gev}, while Ref.~\cite{Boiarska:2021yho} made similar claims for KM3NeT/ORCA.
However, as we will discuss in this article, this rate does not readily translate to constraints at this mixing level due to the large number of background events and particle identification errors. 

\begin{figure}[h]
    \includegraphics[width=\columnwidth]{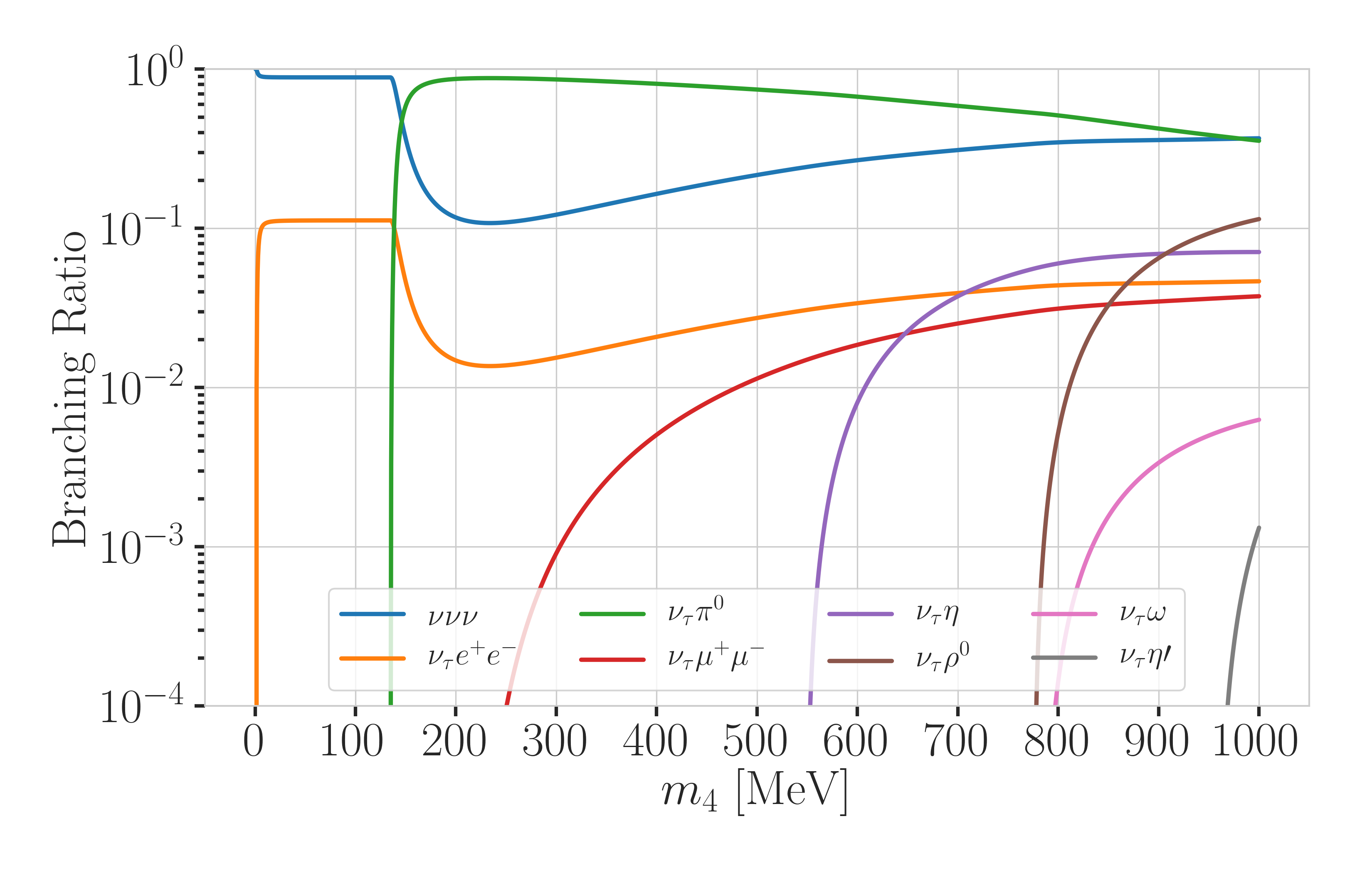}
    \caption{Branching ratios of an HNL within the mass range considered in this work, only considering $\nomathut4 \neq 0$, calculated based on the results from Ref.~\cite{Coloma:2020lgy}.}
    \label{fig:hnl_branching_ratios}
\end{figure}

An HNL interaction in IceCube produces observable light through two processes.
Light is emitted first when the HNL is produced in a hard neutrino scattering process, also producing a hadronic shower, and second when the HNL decays into SM products.
The amount of light emitted in the first process is governed by the inelasticity of the neutrino interaction and light produced in the hadronic shower, while the amount of light emitted in the second process depends on the decay channel.
Since hadronic showers cannot be resolved as such in IceCube, they are observed as cascade-like light depositions. \cref{fig:hnl_branching_ratios} shows the branching ratios of an HNL with mass $m_4$ into SM products. They are derived from the decay widths calculated based on the results from Ref.~\cite{Coloma:2020lgy}, only considering the decays possible through mixing with $\nu_\tau$.
The majority of these decay modes produce a cascade-like light deposition in IceCube, where just the decay with $\mu$'s in the final state can produce elongated light patterns if sufficient energy is transferred to the outgoing $\mu$. 
The dominant decay mode for all HNL masses above the pion production threshold and below \SI{1}{\gev}, the decay to the neutral pion - $\nu_4 \to \nu_\tau + \pi^0$ - produces a pure EM cascade, for example.

The light emissions of the first and second processes, together with the potentially large separation between them, can produce a signature composed of two cascade-like light depositions, therefore called double-cascade signature.
The distance between the two cascades depends on the lifetime of the HNL, which follows an exponential distribution and is strongly dependent on $m_4$ and \ut4, which are the free model parameters.
Adding this morphological signature of two cascades to the known categories in IceCube makes three distinct morphological categories, which are theoretically observable: cascades, tracks, and double-cascades.
However, given the sparse arrangement of the detector modules and the energy detection threshold of a few \si{\gev}~\cite{IceCube:2022kff} for events happening within the proximity of the detection modules ($\mathcal{O}(\si{\meter})$), most events of all three categories only deposit light in one location of the detector and are therefore observed as single cascades.
This means that the majority of HNL events, which theoretically produce double-cascades, actually appear in the atmospheric neutrino sample as an excess of events looking like cascades.
They effectively appear as a shift in the energy and zenith distribution of the cascade component of the sample, accompanied by a small subset of events with a signature composed of two cascades. This distinctive signature is unique at low energies (in the 10s of GeV), and will from here on be referred to as \textit{low-energy double-cascade} signature.

An essential part of this work was the development of an event generator, which was extensively used to study the signal of HNLs in IceCube, crucially factoring in the realistic detector response.
The generator~\cite{LeptonInjector-HNL} is published as an open-source tool to produce HNL events for atmospheric neutrino detectors. 
Based on the generated events, the performance of detecting and identifying HNLs via the double-cascade channel was investigated, but despite this morphology being unique to HNLs, it is not currently possible to observe and correctly identify it, given the large background of SM neutrino events.
As an alternative avenue and serving as a \textit{proof-of-principle} search for HNLs in IceCube, the excess of cascade-like events stemming from HNLs, on top of the expectation from SM interactions, is used to perform a first measurement of \ut4 for three masses with ten years of data from the IceCube DeepCore sub-array~\cite{IceCube:2011ucd}.

%%%%%%%%%%%%%%%%%%%% IceCube DeepCore Detector %%%%%%%%%%%%%%%%%%%%

\section{\label{sec:detector}IceCube DeepCore Detector}

\begin{figure}[t!]
    \includegraphics[width=\columnwidth]{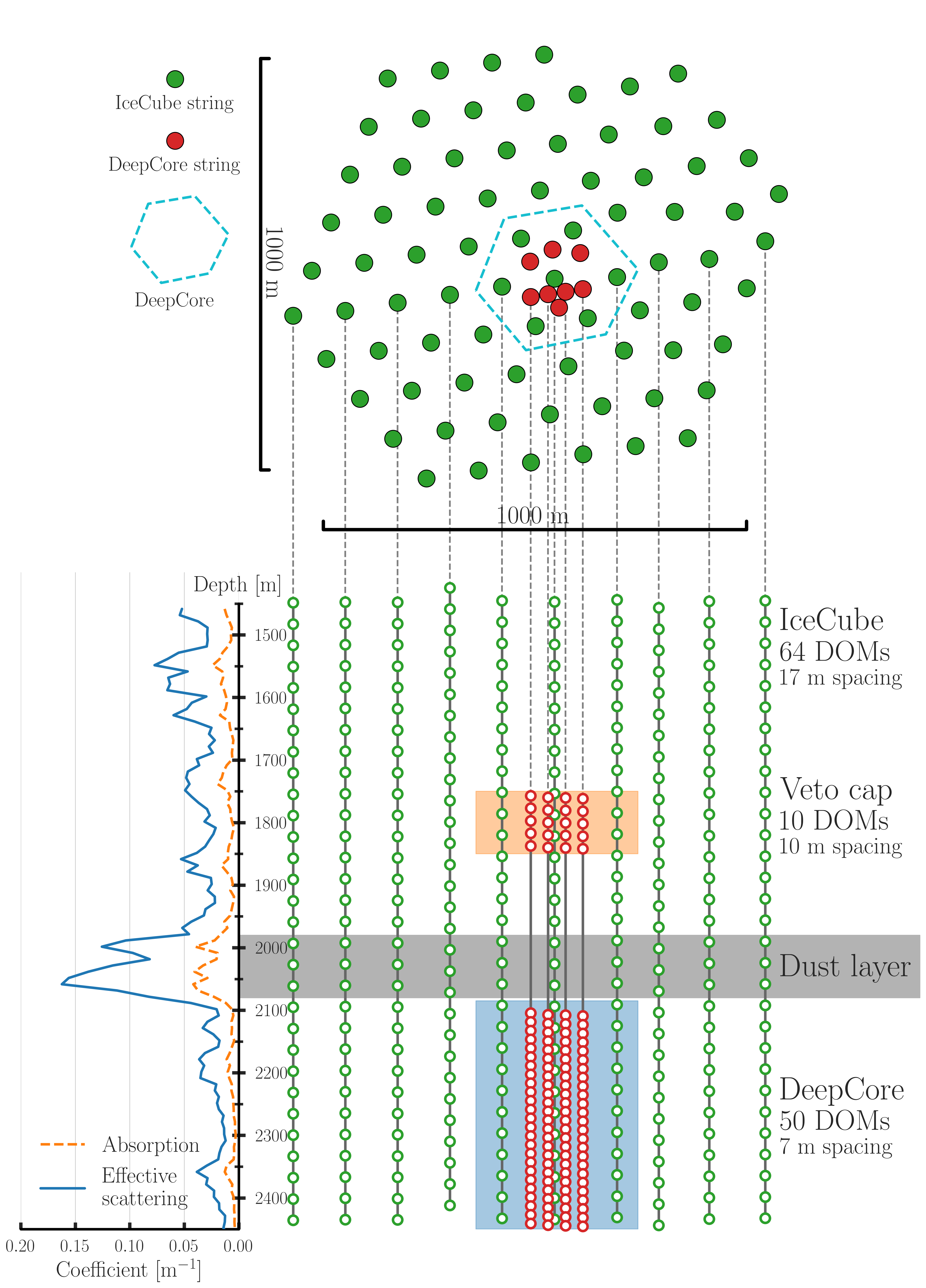}
    \caption{Schematic overview of the IceCube and DeepCore in-ice detector arrays. The denser horizontal and vertical module spacing of the DeepCore sub-array is visible, which lowers the energy detection threshold compared to the sparser IceCube array. Also shown are the effective scattering and absorption coefficients with respect to depth, where the region of significantly inferior optical properties is highlighted and labeled as the \textit{dust layer}~\cite{ice_calibration}. The majority of DeepCore DOMs are located below the dust layer, in the region of ice with the best optical properties.}
    \label{fig:detector_geometry}
\end{figure}

The IceCube detector consists of 5,160 digital optical modules (DOMs) embedded inside the antarctic ice at the South Pole between depths of \SI{1450}{\m} and \SI{2450}{\m}. Each DOM contains a 10" photomultiplier tube (PMT) and associated electronics for high voltage generation, signal capture, and calibration, which are enclosed in a glass pressure vessel. DOMs are connected to vertical cables that provide power and communication via infrastructure on the surface. They are arranged in a roughly hexagonal array, which can be seen in the upper part of \cref{fig:detector_geometry}, while the bottom part shows the vertical distribution of the DOMs in the detector. The DOMs in the main IceCube array (green circles) are separated by a \SI{17}{\m} vertical distance along the cable, with $\sim$\SI{125}{\m} horizontal spacing between the cables. This yields an energy detection threshold of around \SI{100}{\gev}~\cite{IceCube:2016zyt}.

The DeepCore sub-array (red circles) - the more densely instrumented region in the bottom center of IceCube - is optimized to detect neutrinos with energies down to a few GeV. This is achieved by instrumenting the ice at depths between \SI{2100}{\m} and \SI{2450}{\m} where the scattering and absorption lengths of the ice are longest\footnote{A few DeepCore DOMs are located at depths between \SIrange{1950}{1850}{\m} with a \SI{10}{\m} vertical spacing to aid in atmospheric $\mu$ tagging.}, allowing for more efficient photon transport~\cite{ice_calibration}. The ice properties are also shown on the left side panel of \cref{fig:detector_geometry}, where the region of clear ice below \SI{2100}{\m} is visible. DeepCore DOMs have a separation of \SI{7}{\m} vertically and \SIrange{40}{80}{\m} horizontally, and contain PMTs with approximately \SI{35}{\percent} higher quantum efficiency~\cite{IceCube:2011ucd}. Altogether, these features enable the detection of neutrinos down to approximately \SI{5}{\gev}.

%%%%%%%%%%%%%%%%%%%% Heavy Neutral Lepton Simulation %%%%%%%%%%%%%%%%%%%%

\section{\label{sec:generator}Heavy Neutral Lepton Simulation}

A dedicated Monte Carlo generator was developed to simulate the signal events sought after in this analysis. It is an adaptation of the \textsc{LeptonInjector} (LI) package~\cite{IceCube:2020tcq} named \textsc{LeptonInjector-HNL} (LI-HNL)~\cite{LeptonInjector-HNL}. LI is a standalone event generator capable of producing neutrino interactions optimized for the needs of neutrino observatories. LI-HNL makes use of the existing structure of LI, modified to support the HNL's beyond Standard Model (BSM) physics.
This section outlines LI's basic functionality, listing the changes made to support HNL production, and describes the event generation procedure, including a detailed description of the theoretical signature.

\subsection{Overview of LeptonInjector}

LI is a generator specifically designed for neutrino interactions at coarsely instrumented, large-volume neutrino detectors.
Because of this focus, it differs from other common event generators in a few ways.
First, it's optimized for higher energy interactions (above \SI{10}{\gev}), where DIS becomes the dominant neutrino-nucleus scattering process.
Second, it handles neutrino transport through Earth and the neutrino flux separately from event generation.
LI simulates events directly in or around the detector, where for this work, the \textit{volume} injection mode is used.
In this mode, the neutrino interactions are sampled uniformly throughout a chosen volume, usually encompassing the instrumented volume.
This is ideal for interactions where the daughters do not propagate over long distances.
% Both modes increase simulation efficiency by preferentially simulating those events which can trigger the detector.
% , with either a \textit{ranged} or \textit{volume} injection mode.
% Ranged mode is designed for interactions where the daughter particles themselves travel long distances before being detected --~e.g., $\nu_\mu$ interactions, where the daughters of interest are $\mu$'s.
% In ranged mode, the neutrino interaction is simulated at a distance from the detector and out-going charged leptons are propagated to the detector~\cite{IceCube:2020tcq}.
LI directly injects the final states of neutrino interactions. For example, in a charged-current (CC) DIS interaction, LI generates a lepton and a hadronic shower at the interaction vertex of the neutrino and assigns the relevant final state kinematics by sampling from the corresponding differential cross-sections.

Neutrino flux and transport through Earth are handled by a complementary package, \textsc{LeptonWeighter} (LW)~\cite{IceCube:2020tcq} with inputs from \textsc{MCEq}~\cite{Fedynitch:2018cbl} and \textsc{nuSQuIDS}~\cite{Arguelles:2021twb}.
LW assigns a probability weight to each event accounting for the likelihood it is produced given the user's chosen flux model, neutrino oscillations through Earth, and neutrino cross-section model.
These weights can then be used to calculate event rates for a particular detector and event selection.

\subsection{Customizations of LeptonInjector-HNL}

LI-HNL modifies LI to support the production of HNLs from $\nu_\tau$-neutral-current (NC) interactions. 
The lepton produced at the interaction vertex is chosen to be the HNL or anti-HNL. 
After a chosen distance, the HNL is forced to decay, and a set of secondary particles are generated at the decay point. The decay also happens via a NC weak interaction.
The distance is chosen from an ad-hoc power-law distribution, with minimum and maximum values that maximize the selection efficiency in this analysis. The physically correct distribution is obtained through the re-weighting process described in \cref{sec:weighting}.
These parameters are customizable for other detectors' needs.
The generator returns an event tree containing information about the injected $\nu_\tau$, the resulting HNL and the hadronic shower, and the information for each of the decay daughter particles of the HNL.
The addition of these secondary daughters is a key functionality of LI-HNL compared to LI.

The mass of the HNL is a free parameter in the model under investigation.
% As motivated in the $\nu$ Minimal Standard Model introduced in Refs.~\cite{ASAKA200517, ASAKA2005151}, extending the SM by three right-handed neutrinos - where two of them are heavy and one of them is light - could explain both the origin of neutrinos masses, while also providing a dark matter candidate. This work will therefore focus on HNLs in the GeV regime.
% , where modifications of the see-saw mechanism can account for light neutrino masses without introducing complications to the hierarchy problem, as discussed in Refs.~\cite{Vissani:1997ys,Coloma:2017ppo}.
Because the mass affects the production and decay kinematics, as well as the accessible decay modes, multiple discrete mass samples are tested individually.
For the analysis described in this work, three samples are produced for masses of \SI{0.3}{\gev}, \SI{0.6}{\gev}, and \SI{1.0}{\gev}, which are the supported masses of the generator. It additionally supports an HNL mass of \SI{0.1}{\gev}, but no HNL masses above \SI{1}{\gev}, because above \SI{1}{\gev} decays to multiple hadrons becomes possible, that pose unique simulation challenges beyond the scope of this work

% For HNL masses above 1 GeV, a decay to additional hadrons is opened, which poses unique simulation challenges beyond the scope of this work.

\begin{figure}[h]
    \includegraphics[width=0.9\columnwidth, trim=0 3.5cm 0 0, clip]{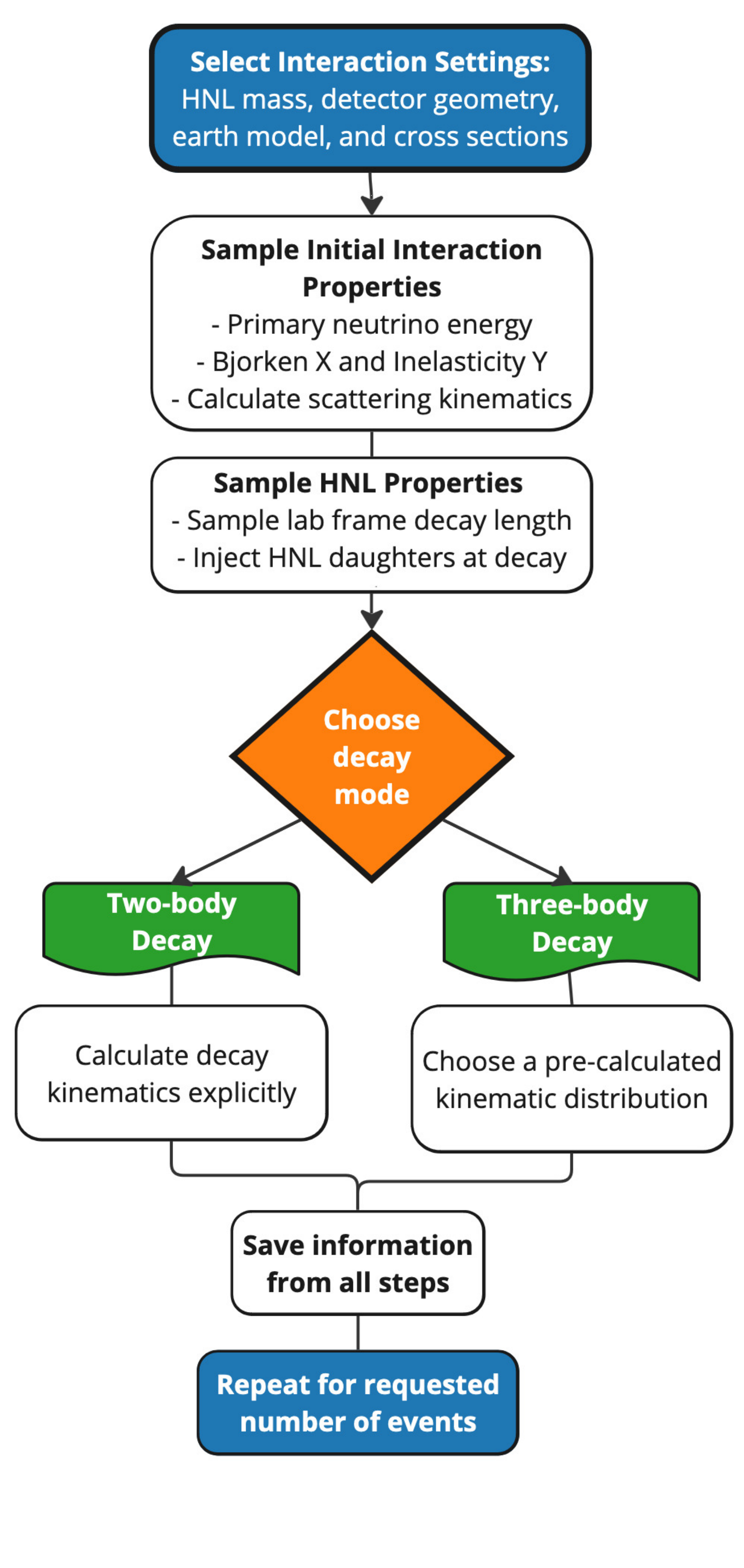}
    \caption{Event generation procedure in \textsc{LeptonInjector-HNL}.
    Bjorken $x$ and inelasticity are selected from distributions shown in \cref{fig:bjorken_x_inelasticity_y}.}
    \label{fig:flowchart}
\end{figure}

\begin{figure}[h]
    \centering
    \includegraphics[width=\columnwidth]{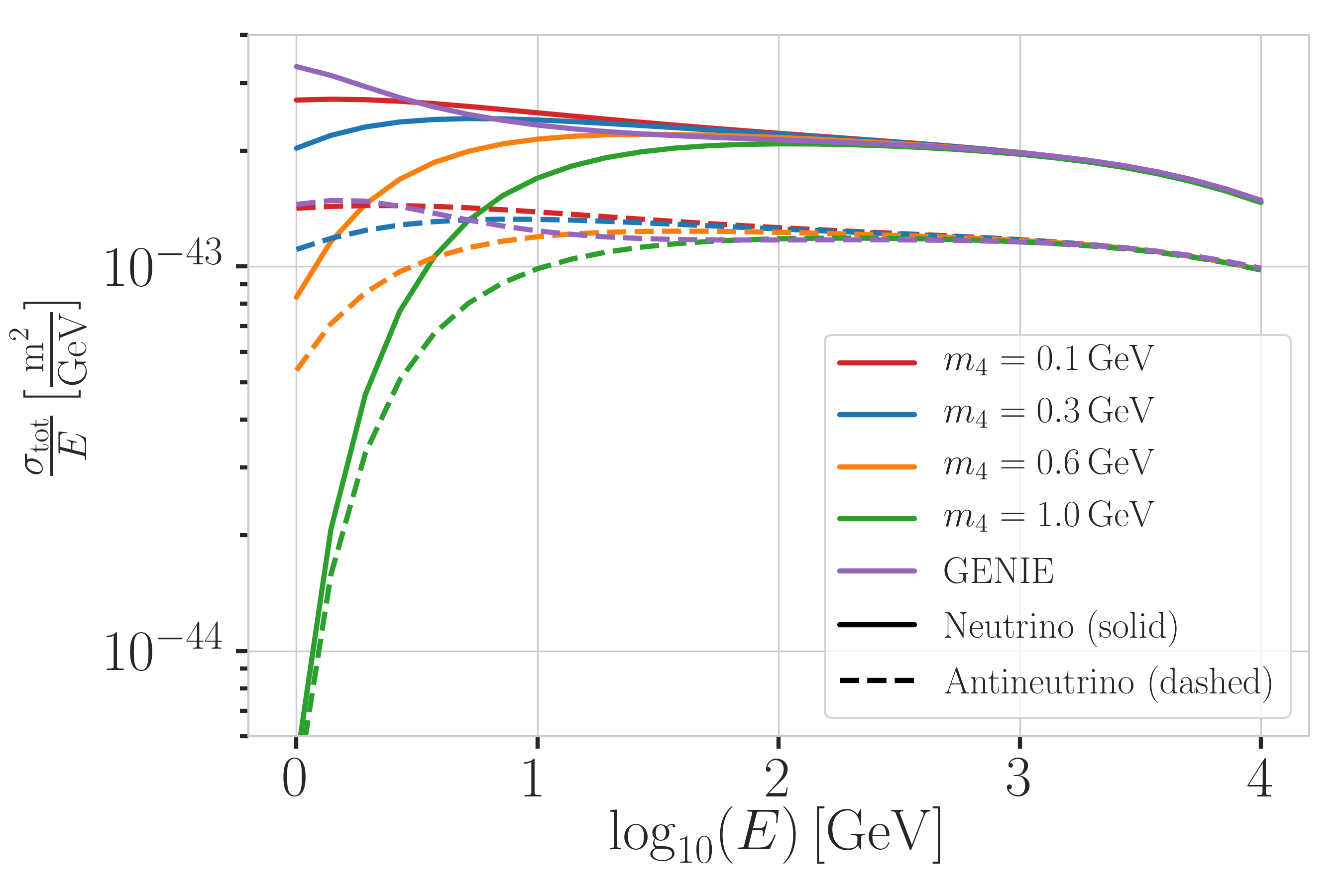}
    \caption{Total neutrino-nucleon cross-sections for the $\nu_\tau$/$\bar{\nu}_\tau$ up-scattering into the heavy mass state, shown for the three target masses, compared to the total SM $\nu_\tau$/$\bar{\nu}_\tau$-NC cross-sections used for the baseline neutrino simulation production with \textsc{GENIE}. The plot shows the cross-sections for a mixing of $\nomathut4=1$, since the cross-sections scale linearly with the mixing \ut4 and the plot aims to visualize the shape compared to the SM case.}
    \label{fig:total_cross_section}
\end{figure}

The generation procedure is illustrated in \cref{fig:flowchart}, highlighting the HNL-specific portions of the generation software. The interaction begins with a sampling on HNL production kinematics from custom cross-sections. These cross-sections were computed using \textsc{NuXSSplMkr}~\cite{xsecmaker}, a tool that calculates neutrino-nucleon cross-sections from parton distribution functions and fits them to a spline surface.
Specifically, LI uses \textsc{Photospline}~\cite{Whitehorn:2013nh} to produce b-splines in a format readable by LI and LW.
For this work, \textsc{NuXSSplMkr} was modified to produce the $\nu_\tau+N \to \nu_4+X$ cross-sections by adding a kinematic condition to the $\nu_\tau$ interaction which ensures there is enough energy to produce the outgoing heavy mass particle, $\nu_4$.
This is the same kinematic condition used in the $\nu_\tau$-CC process, which ensures that sufficient energy is present to produce the massive charged lepton~\cite{Levy:2004rk}.
Compared to the SM cross-section, the additional kinematic condition reduces the cross-section at lower energies, where the energy available to produce the HNL is more often lower than the HNL mass.
This is shown in \cref{fig:total_cross_section}, where the cross-sections generated for this analysis are compared to the SM NC cross-sections calculated using \textsc{GENIE}~\cite{Andreopoulos:2009rq}.

The \texttt{GRV98LO} parton distribution functions~\cite{Gluck:1998xa} are used for cross-section calculations in this work.
As expected, the cross-sections produced agree with the GENIE cross-section above $\approx \SI{200}{\gev}$. This modification is available as part of the \textsc{NuXSSplMkr} package~\cite{xsecmaker}.
In general, LI-HNL supports simulation with any total or double-differential cross-sections, so long as splines have been generated using \textsc{Photospline}.

\begin{figure}[h]
    \includegraphics[width=\columnwidth]{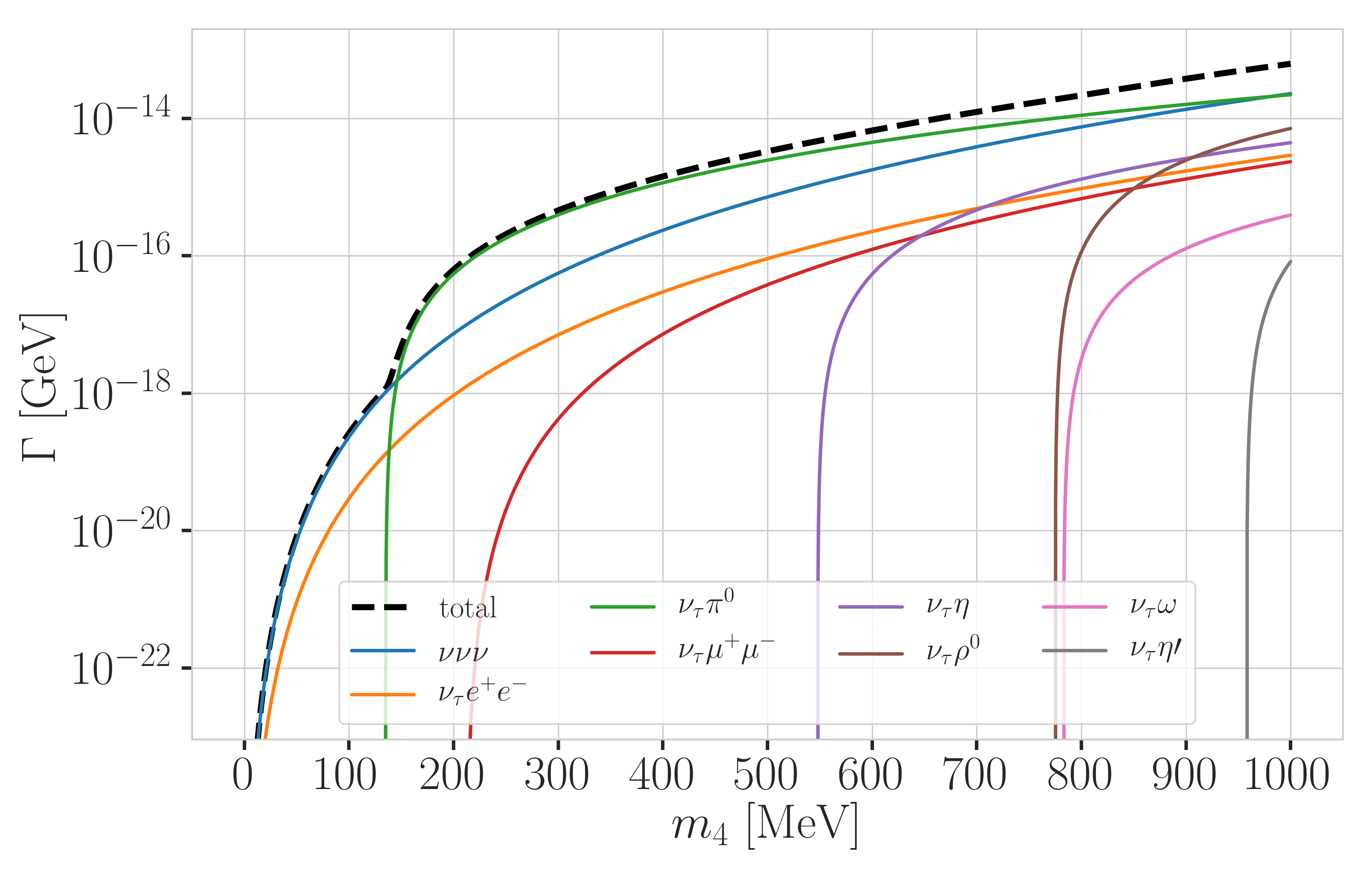}
    \caption{Shown are decay widths of the HNL within the mass range considered in this work, assuming mixing only with the $\tau$ sector. As the decay widths scale linearly with the mixing parameter \ut4, this plot simply uses $\nomathut4=1$ to visualize their shape.
    % Since the decay widths scale linearly with the mixing, it can be factored out and the plot shows the decay widths for a mixing of $\nomathut4=1$ to visualize their shape.
    Calculated based on the results from Ref.~\cite{Coloma:2020lgy}.}
    \label{fig:hnl_decay_modes_log_decay_width}
\end{figure}

The HNL decay daughters are created after the sampled decay length, where the decay mode is selected according to the branching ratios of the corresponding HNL mass shown above in \cref{fig:hnl_branching_ratios}.
% The branching ratios are derived from the decay widths calculated based on the results from Ref.~\cite{Coloma:2020lgy}, only considering the decays possible through mixing with $\nu_\tau$.
The total decay width and its individual components as a function of the HNL mass are shown in \cref{fig:hnl_decay_modes_log_decay_width}. 

\begin{figure}[h]
    \centering
    \includegraphics[width=\columnwidth]{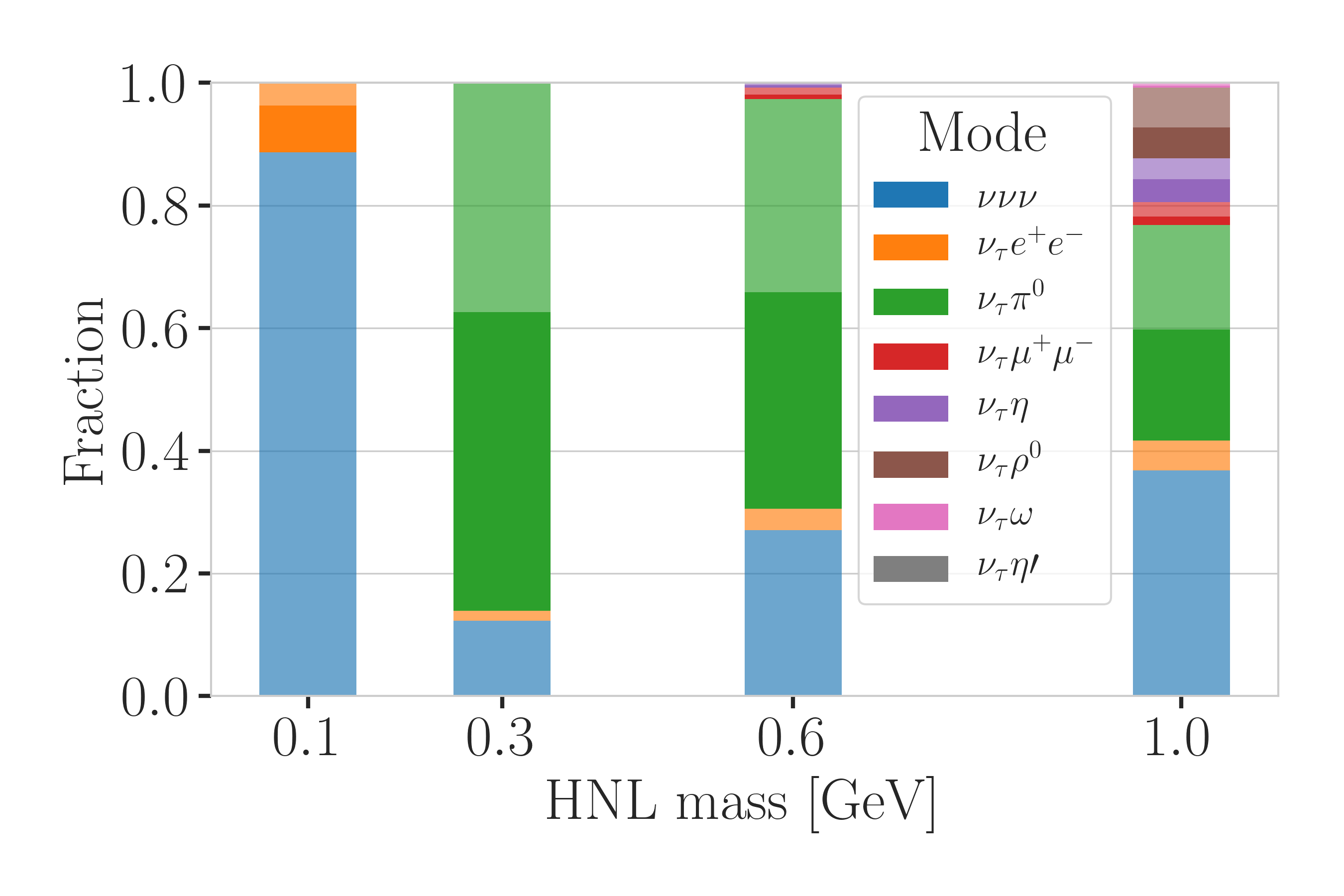}
    \caption{Decay modes for HNL masses considered in this work. The colored bars show the branching ratios of the different decay modes, where the dark and light portions are showing the median visible and invisible energy fraction of the particular decay mode, respectively. The three-neutrino mode, for example, is completely invisible, and thus shown only as the lighter shade.}
    \label{fig:hnl_branching_ratios_data}
\end{figure}

The decay mode strongly influences the amount of potentially visible energy produced by the decay. In particular, the kinematics of each decay determine how much energy goes into the non-neutrino particle (and therefore can be visible) compared to the neutrino (and is therefore invisible). The compositions of decay types at four selected mass values are shown in \cref{fig:hnl_branching_ratios_data}. The chart shows each decay mode as a colored portion of the bar, where the more saturated portion represents the fraction of the decay's energy which is deposited as visible light in the detector. 
Note that, for example, the three-neutrino decay mode - which is dominant for an HNL with a mass of 0.1 GeV - is completely unsaturated, resulting in a strongly reduced amount of detectable energy in the detector. For this reason, HNLs with masses of approximately  \SI{0.1}{\gev} were not pursued in this analysis.
% Note, for example, that the three-neutrino decay mode is completely unsaturated, and that this decay mode is dominant for an HNL with a mass of \SI{0.1}{\gev}, resulting in a strongly reduced amount of detectable energy in the detector. This one of the reasons why the \SI{0.1}{\gev} case was not pursued in the final analysis.

Once a decay mode has been selected, the daughter particles are produced accordingly. For two-body decays, the trivial kinematics are calculated directly in LI-HNL. In the case of three-body decays, the kinematic properties are uniformly selected from a list of pre-generated kinematic distributions. These distributions were calculated for each decay mode and at each selected HNL mass using \textsc{MadGraph5}\footnote{Due to the precision of the \textsc{MadGraph} simulation, this introduces an error of order $\SI{e-6}{\gev}$ in the kinematic distributions of the HNL decay products. This imprecision is significantly smaller than the leading sources of uncertainty and does not impact the results of this analysis.} version 3.1.4~\cite{Alwall:2014hca}.
We have used the HNL model described by the \textsc{FeynRules} package, which extends the SM by the addition of a GeV-scale HNL, making them heavy enough to decay to mesons in the final state~\cite{Coloma:2017ppo}.
The decay products are calculated in the rest frame of the HNL, and are then boosted to the lab frame.
% The use of pre-computed kinematic tables eliminates the runtime computational load of calculating complex three-body decay kinematics.

\subsection{Event Generation}

To simulate events, the LI settings have to be specified, which includes selecting the injection mode (ranged or volume), the Earth's density profile, the cross-section model, and the initial and final particle types.
For this analysis, LI-HNL is run in volume mode, and two generators are used to produce half of the events from $\nu_\tau$ and the other half from $\bar{\nu}_\tau$ as the primary particle. Volume mode injects the primary particles within a cylindrical volume, where the center and the dimension are chosen at runtime. Here, the volume is chosen to be centered in the DeepCore sub-array, and we produce samples with masses of \SI{0.3}{\gev}, \SI{0.6}{\gev}, and \SI{1.0}{\gev}.
Both the production and the decay of the HNL are presumed to occur inside the detector, which is achieved by placing the primary interaction point inside the detector and limiting the decay length range to be below the size of the detector, which is $\sim$\SI{1000}{\meter}. The parameters chosen for the sample generation are listed in \cref{tab:sampling_dists}. 

\begin{table}[h]
    \small
    \begin{tabular}{ lll }
    \hline\hline    
    \textbf{Variable} & \textbf{Distribution} & \textbf{Range/Value} \\     
    \hline\hline    
    energy & $E^{-2}$  & [$\SI{2}{\gev}$, $\SI{10000}{\gev}$] \\
    zenith & uniform in $\cos{\theta}$ & [$80^{\degree}$, $180^{\degree}$] \\
    azimuth& uniform & [$0^{\degree}$, $360^{\degree}$] \\
    vertex \textit{x, y}& uniform on circle & $r=600$ m \\
    vertex \textit{z} & uniform  & - 600 m  to 0 m \\
    $m_4$& fixed & {0.3, 0.6, 1.0} GeV \\
    $L_{\textrm{decay}}$ & $L^{-1}$ & [0.0004, 1000] m \\
    \hline
\end{tabular}
\caption{Analytical sampling distributions used in LI-HNL for this analysis.}
\label{tab:sampling_dists}
\end{table}

\begin{figure}[h]
    \centering
    \includegraphics[width=0.9\columnwidth]{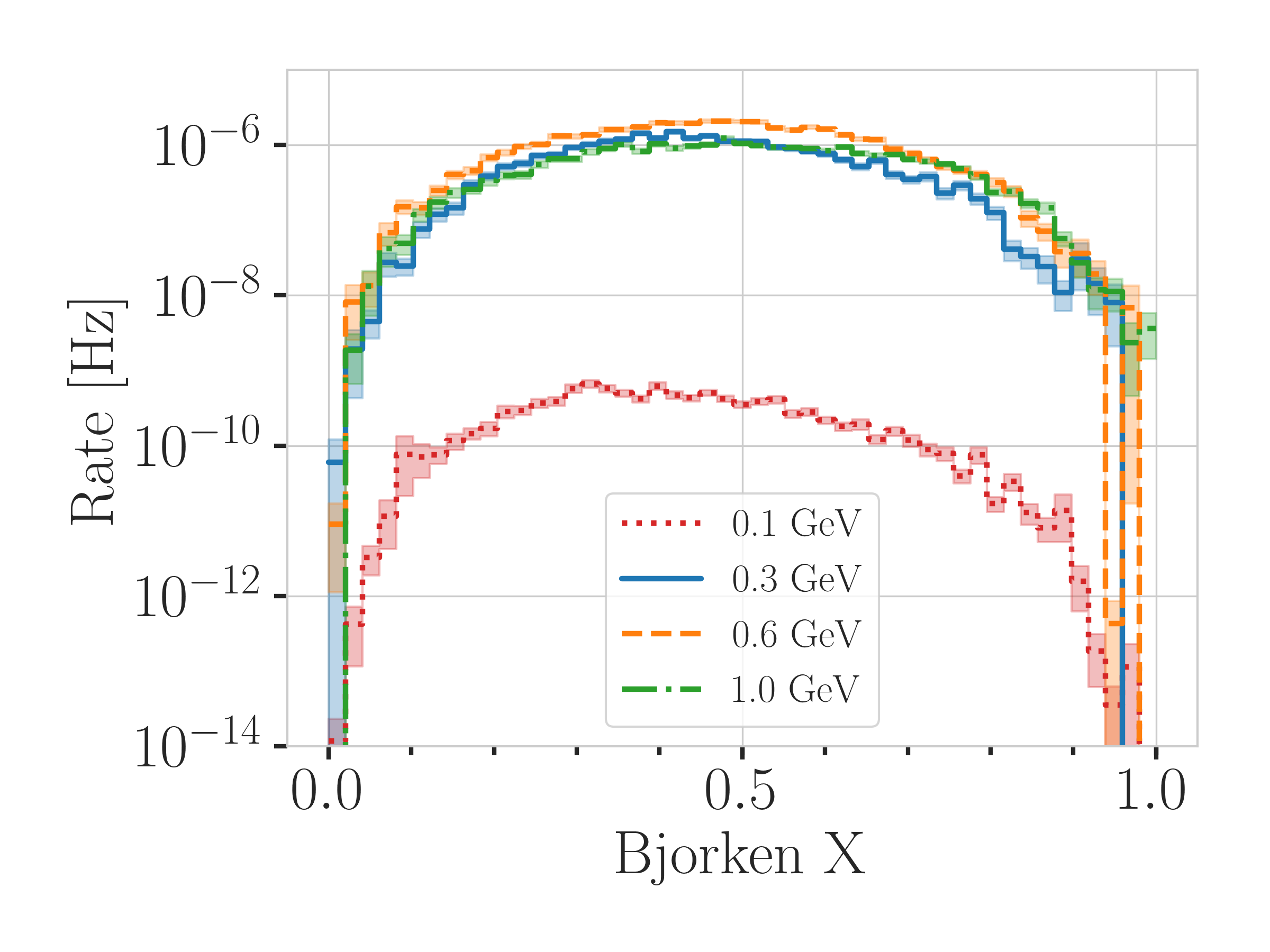}
    \includegraphics[width=0.9\columnwidth]{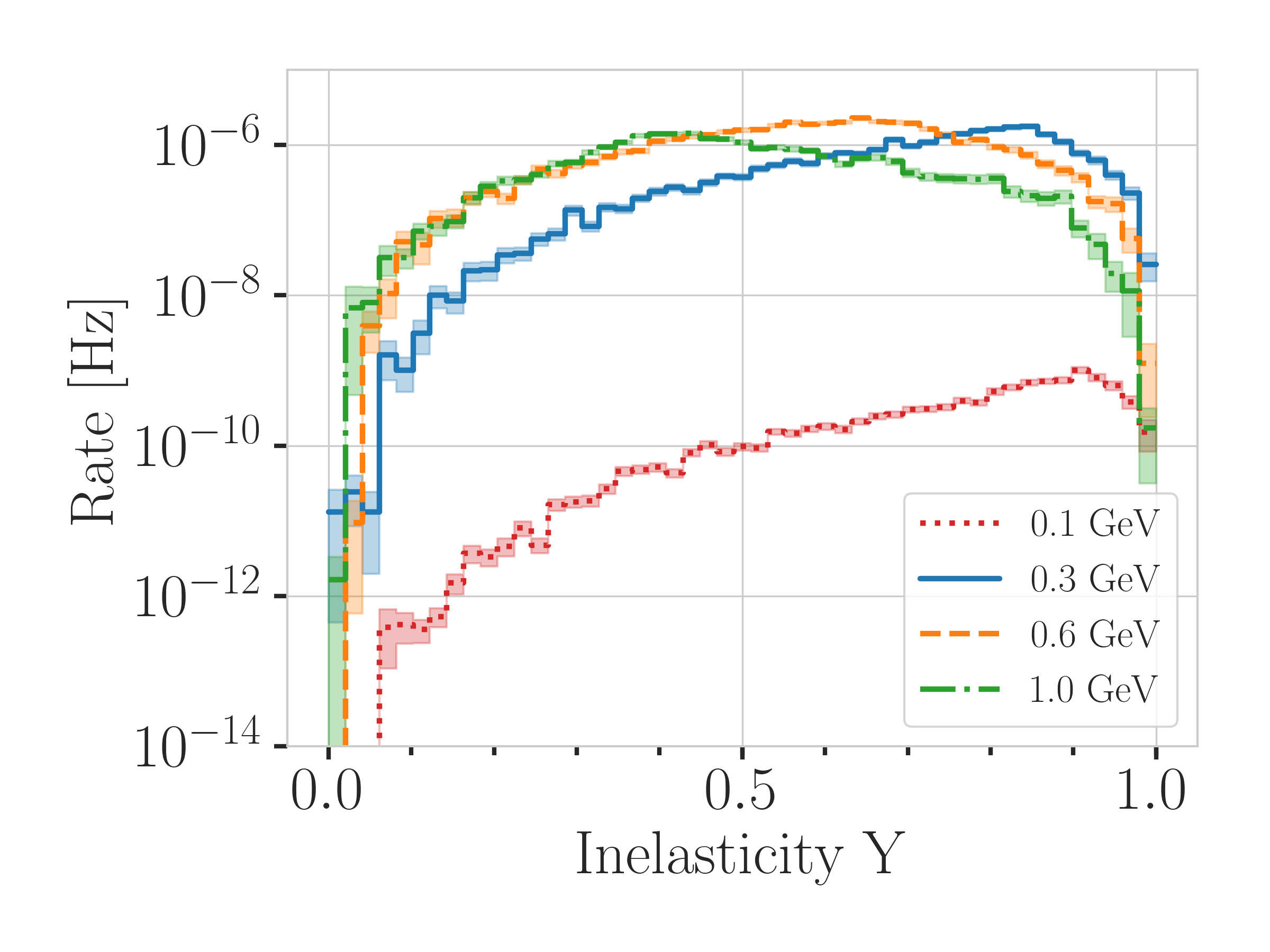}
    \caption{Generation level Bjorken $x$ distribution (top) and inelasticity $y$ distribution (bottom) for the three mass samples. Shown are the distributions from 25000 MC events, weighted according to the baseline atmospheric flux and oscillation parameters and to a mixing of $\nomathut4 = 0.001$. The shaded bands indicate the statistical uncertainty.}
    \label{fig:bjorken_x_inelasticity_y}
\end{figure}

\cref{fig:bjorken_x_inelasticity_y} shows the generation level distributions of Bjorken $x$ and $y$\footnote{The $y$ distributions used for the analysis looks slightly different, since during post-unblinding checks, a bug was discovered in the spline-fitting procedure of the differential cross-sections that led to inconsistencies in the generated inelasticity distribution. The bug has been fixed in both \textsc{NuXSSplMkr} and LI-HNL, and was shown to have a negligible impact on the final analysis results.} used for the HNL production. These distributions are the result of sampling from the custom HNL cross-sections that depend on energy as well as $x$ and $y$. The Bjorken variables have a direct influence on the fraction of visible energy in each event. Visible energy is defined here as the energy contained in the hadronic shower and non-neutrino decay products, which can produce Cherenkov radiation visible in the IceCube detector.
As a demonstrative example, the distributions are weighted to a mixing of $\nomathut4 = 0.001$, which leads to a reduced expected rate for the \SI{1.0}{\gev} case, because the expected decay lengths become larger than the chosen simulation range (compare the example behavior shown in \cref{fig:hnl_decay_feynman_lengths}).
It can be seen how a larger HNL mass leads to the Bjorken $x$ distribution peaking at larger values, because the momentum transfer increases. This on the other hand leads to a reduction in the inelasticity $y$, which can be seen by the $y$ distribution peaking towards smaller values for increasing HNL mass.

\begin{figure}[h]
    \centering
    \includegraphics[width=0.9\columnwidth]{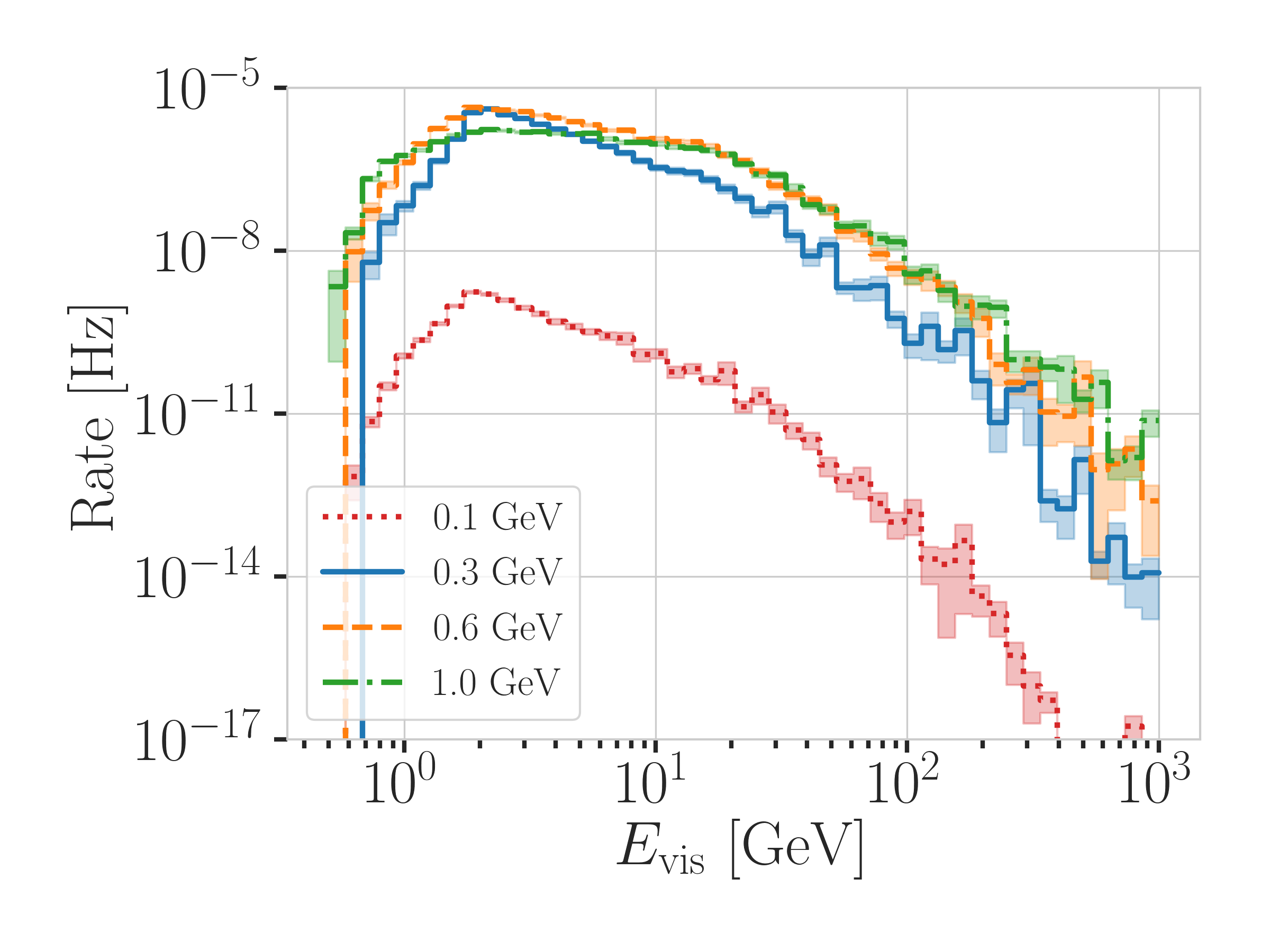}
    \includegraphics[width=0.9\columnwidth]{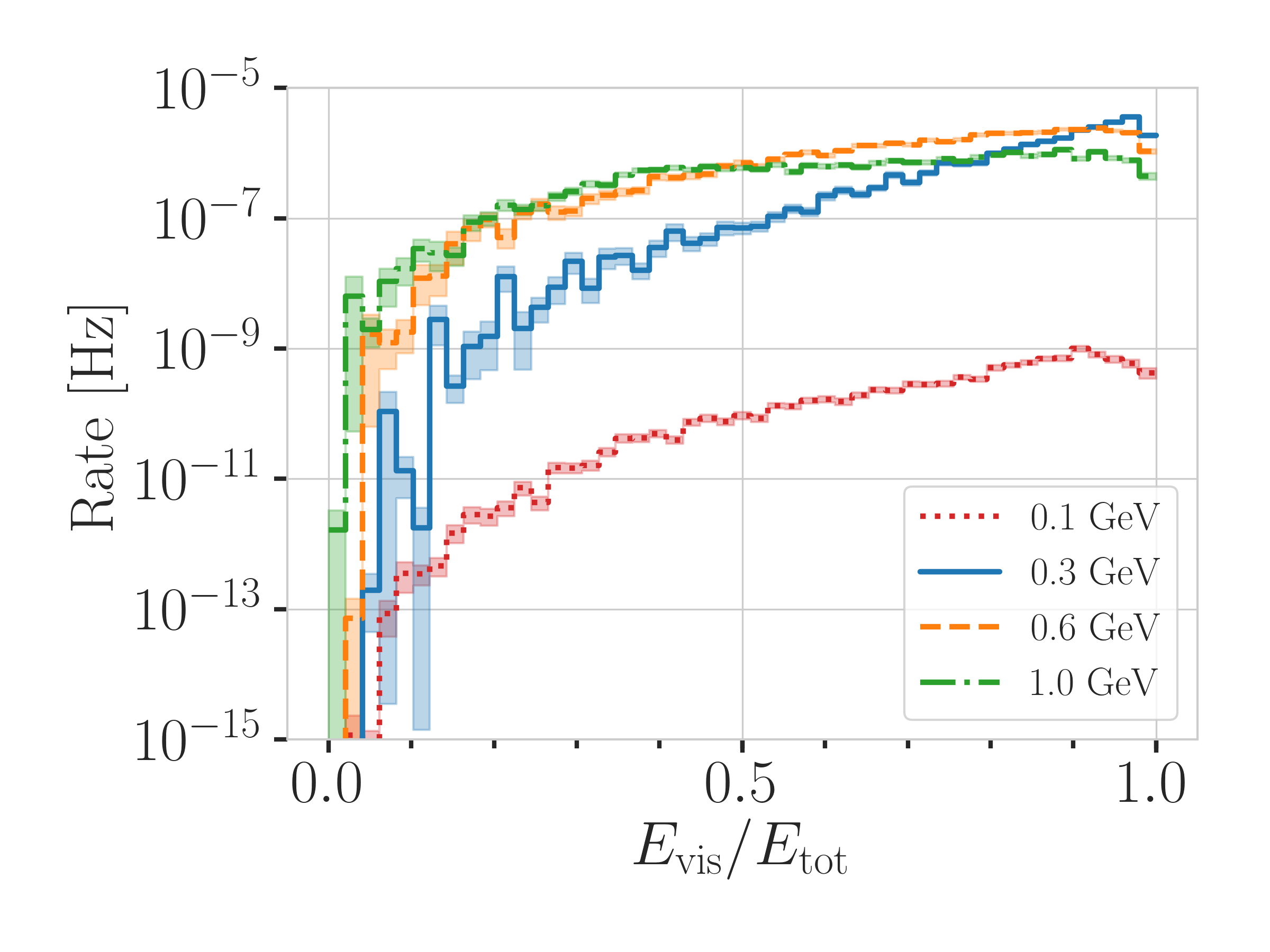}
    \caption{Generation level total visible energy distribution (top) and visible energy fraction with respect to the total true neutrino energy (bottom). Shown are the distributions from 25000 MC events, weighted according to the baseline atmospheric flux and oscillation parameters and to a mixing of $\nomathut4 = 0.001$. The shaded bands indicate the statistical uncertainty.    
    }
    \label{fig:visible_energies}
\end{figure}

\cref{fig:visible_energies} shows the visible energy distribution and the distribution of the visible energy fraction, again weighted to a mixing of $\nomathut4 = 0.001$, chosen as a demonstrative example.
The peak in the visible energy fraction of the \SI{1.0}{\gev} case around a fraction of 0.4 is a result of the inelasticity distribution peaking at lower values combined with a large proportion of decays into three neutrinos, resulting in only the energy from the first cascade being visible.
All visible energy fraction distributions are increasing towards 1.0, but the visible energy of the \SI{0.3}{\gev} case peaks strongest towards 1.0, because its main decay is the $\pi^0$ mode and the invisible three-neutrino decay is subdominant. These plots show that already at generation level, a large portion of the energy is invisible to the detector.

% TODO: Include if the package will go public with the paper, otherwise not really necessary information.
% Kinematic checks are implemented throughout the generator, to ensure all selected properties are physically compatible with one another. The entire simulation procedure has been validated through a suite of unit tests included in the LI-HNL package.

\subsection{Detector Simulation}

After generation, the HNL events are subsequently processed in a similar fashion to neutrino events being treated in Ref.~\cite{IceCubeCollaboration:2023wtb}. The hadronic shower from initial interaction is simulated using \textsc{GEANT4}~\cite{GEANT4:2002zbu} for energies below \SI{30}{\gev}, and using an internal cascade simulation package for higher energies. It computes the Cherenkov yield from analytical approximations derived in Ref.~\cite{Radel:2012kw}.
The decay products of the HNL are simulated using various packages depending on the particle produced.
\textsc{Proposal}~\cite{Koehne:2013gpa} is used to propagate $\mu$'s and to computes their Cherenkov light output.
The shower development of gamma rays, electrons, and positrons below \SI{100}{\mev} is simulated using \textsc{Geant4}, while at higher energies, the internal cascade simulation package is used.

After event generation and particle propagation, two simulation steps remain: the propagation of photons and simulation of detector response.
Photon propagation is performed by \textsc{clsim}~\cite{clsim}, an \textsc{OpenCL} implementation that yields equivalent results as the Photon Propagation Code (\textsc{PPC})~\cite{Chirkin:2019rcj}.
The ice properties relevant for photon propagation are described by the South Pole Ice (SPICE) model~\cite{IceCube:2013llx}, which accounts for inhomogeneities in the ice, as well as tilting due to the uneven bedrock below the glacier.
Finally, the detector response includes a characterization of the total efficiency and angular acceptance of each DOM.
The total efficiency takes into account the DOM glass transmission probability and the PMT quantum and photo-electron collection efficiencies as a function of wavelength~\cite{ABBASI2009294_data_acquisition}.
The detector simulation also incorporates corrections for local ice effects, due to the distinct nature of refrozen ice in the bore-holes in which the DOMs are situated, and photons created by radioactive decays in the DOMs themselves.
This part of the simulation is handled by IceCube proprietary software.

\subsection{Monte Carlo Re-Weighting} \label{sec:weighting}

\begin{figure}[h]
    \includegraphics[width=0.9\columnwidth]{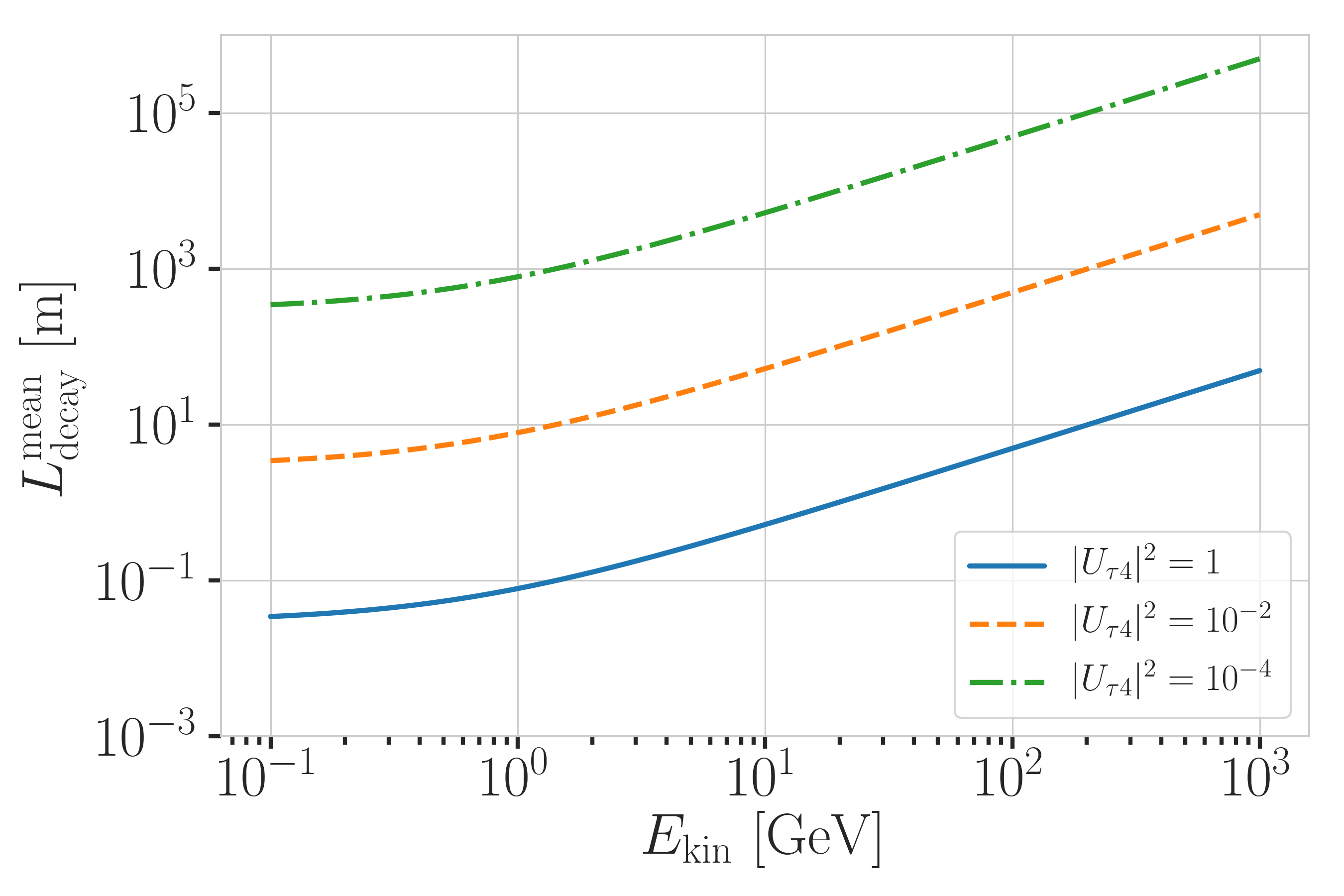}
    \caption{Mean decay length of the HNL for a mass of \SI{0.6}{\gev} shown for different mixing values.}
    \label{fig:hnl_decay_feynman_lengths}
\end{figure}

The weighting procedure mainly follows the LI workflow, using the \textsc{LeptonWeighter} package, with additional functionality to account for the physical probability of HNL events in the detector.
The user-specified cross-section and flux files are loaded and used to calculate a flux weight (taking into account both the atmospheric neutrino flux and their oscillations using \textsc{NuSQUIDS}) and a LI weight (accounting for the sampling distributions and earth model used in the base LI package).
This analysis adds another weight which transforms the ad-hoc $1/L$ distribution, used to sample the HNL decay length, $L$, into the physically expected exponential distribution.
This weight accounts for the dependence of the decay length on the considered value of \ut4 and the mass $m_4$, where the mean lab frame decay length, defining the exponential distribution, is given by
\begin{equation}
    L_\mathrm{decay} = \gamma v \tau_\mathrm{proper}
    \;,
    \label{eq:decay_length}
\end{equation}
with $\gamma$ being the Lorentz factor of the HNL, $v$ the speed, and the proper lifetime
\begin{equation}
    \tau_\mathrm{proper} = \frac{\hbar}{\Gamma_\mathrm{total}(m_4, \nomathut4)} = \frac{\hbar}{\Gamma^*_\mathrm{total}(m_4) \cdot \nomathut4}
    \;.
    \label{eq:proper_lifetime}
\end{equation}
Here $\hbar$ is the reduced Planck constant, $\Gamma_\mathrm{total}(m_4, \nomathut4)$ the total decay width for a specific HNL mass and mixing strength, \ut4, while $\Gamma^*_\mathrm{total}(m_4)$ is the same quantity with the mixing factored out as shown in \cref{fig:hnl_decay_modes_log_decay_width}.
\cref{fig:hnl_decay_feynman_lengths} illustrates the mean decay length for the example mass of $m_4=\SI{0.6}{\gev}$ and some selected mixing values as a function of the kinetic energy of the HNL.
Applying all weights transforms the generated event distributions into a physical interaction rate in the detector.

%%%%%%%%%%%%%%%%%%%% Event Filtering and Reconstruction %%%%%%%%%%%%%%%%%%%%

\section{\label{sec:filter_and_reco}Event Filtering and Reconstruction}

In this section, we first describe the filtering procedure, briefly introduce the current standard low-energy DeepCore reconstruction, and then describe first attempts to reconstruct the low-energy double-cascade morphology and their resulting performance.

\subsection{\label{sec:filter}Filtering}

The analysis uses data collected between May 2012 and May 2021, only utilizing physics runs, where the data acquisition was functioning without issues resulting in a total detector live time of $\sim$\SI{3387}{days} or \SI{9.28}{years}. The basic event selection used in this analysis starts with the \textit{DeepCore Common Data Sample}, which is described in Section III of Ref.~\cite{IceCubeCollaboration:2023wtb}, and continues with further selection criteria and machine-learning reconstructions introduced in Ref.~\cite{IceCube:2024xjj}.
The resulting data sample contains 150,257 events with an atmospheric $\mu$ contamination of \SI{0.6}{\percent}, estimated from simulation studies.
Also, at this final level and for the most likely values of the oscillation and systematic parameters found in Ref.~\cite{IceCube:2024xjj}, the sample is dominated by $\nu_\mu$-CC events (\SI{58.8}{\percent}) with sub-dominant contributions of $\nu_e$-CC (\SI{23.5}{\percent}) and $\nu_\tau$-CC (\SI{5.8}{\percent}) interactions.
Notably, \SI{11.3}{\percent} of the sample corresponds to NC interactions, which is relevant for this analysis as this indicates the maximum possible proportion of events from HNL NC up-scattering interactions.
% \cref{fig:selection_efficiency} visualizes how the filtering strongly reduces the atmospheric $\mu$ and noise components, while retaining the neutrino and HNL events.
For further details on the basic event selection used in this analysis we refer the reader to Refs.~\cite{IceCubeCollaboration:2023wtb,IceCube:2024xjj}.

% \begin{figure}[h]
%     \includegraphics[width=\columnwidth]{rates_per_level_hnl.png}
%     \caption{Selection efficiency of the filter levels aiming at reducing atmospheric $\mu$'s and noise. Shown are the rates of all SM backgrounds as well as the HNL rates for the three mass samples at a mixing of $\nomathut4 = 0.1$.}
%     \label{fig:selection_efficiency}
% \end{figure}

\subsection{\label{sec:flercnn} Standard Reconstruction}

The analysis described in this work is based on the results of one of the standard low-energy reconstructions used in IceCube DeepCore, which is also used in the $\nu_{\mu}$ disappearance result presented in Ref.~\cite{IceCube:2024xjj}.
It is a convolutional neural network (CNN)-based method~\cite{Micallef:2022jvd, flercnn_proceedings, flercnn}, optimized to reconstruct IceCube events with energies below \SI{100}{\gev}.
The application of a CNN is well motivated by the approximate translational invariance of event patterns in the IceCube detector, resulting in a fast and efficient method.
Only the central part of the detector array --- the eight DeepCore strings and 19 IceCube strings --- are used, while the outermost strings are excluded.
Due to the different $z$-positions of the DOMs on DeepCore and IceCube strings, they are handled by two sub-networks that are combined in the final steps of the reconstruction.
The horizontal position of the DOMs is not used, since the string pattern is irregular.
Individual DOM information is summarized into five charge and time variables; the total summed charge, the time of the first hit, the charge-weighted mean time of the hits, the time of the last hit, and the charge-weighted standard deviation of the hit times. These five variables are then fed into the CNN in addition to the string number and the vertical DOM position.

Three separate networks are trained to perform regression of the events' starting vertex ($x,y,z$ position), the energy, and the cosine of the zenith angle.
Two more are used to perform classification tasks, predicting the probability of the event to be track-like ($\nu_\mu$-CC) and the probability of the event being an atmospheric $\mu$. Further detail and performance metrics can be found in Refs.~\cite{Micallef:2022jvd, flercnn_proceedings}.

\subsection{\label{sec:taupede} Double-Cascade Reconstruction}

Events from low-energy atmospheric neutrinos either produce a pure cascade ($\nu_{e}$-CC, \SI{83}{\percent} of $\nu_{\tau}$-CC, $\nu$-NC) or a track and cascade ($\nu_{\mu}$-CC, \SI{17}{\percent} of $\nu_{\tau}$-CC) morphology, while the production and the decay of an HNL inside the detector, however, can produce the double-cascade morphology, as explained in \cref{sec:introduction}. The algorithm used to attempt to reconstruct and identify this morphology is based on the reconstruction method used to search for high-energy, astrophysical $\nu_{\tau}$'s~\cite{Abbasi:2020zmr} and was first introduced in Refs.~\cite{PHallen, MUsner}. Like most traditional IceCube reconstructions~\cite{IceCube:2013dkx,IceCube:2024csv}, it uses a maximum likelihood algorithm, comparing the light observed in the detector to the expected light distribution for a given event hypothesis, through a Poisson likelihood.
% Different event hypotheses can be tested, where the parameter set defining the specific event is $\Theta$.
The double-cascade hypothesis is composed of two cascades aligned with the neutrino direction separated by a certain distance. Assuming the HNL is relativistic, this geometry can be defined by nine parameters: the position of the first cascade, $x, y, z$, the direction of the neutrino, $\phi, \theta$, the interaction time, $t$, and the distance from first to second cascade, $L$, and the energy of each cascade, $E_0$ and $E_1$. Thus, the nine parameters defining an event are $x, y, z, t, \theta, \phi, E_0, E_1, L$.

The energy reconstruction was found to perform well for $E_0$ and $E_1$ above true cascade energies of around \SIrange{8}{10}{\gev}. Below this energy, little or even no light is observed in the detector and the reconstruction performs worse or fails. Due to the kinematic effects explained in \cref{sec:generator}, a large fraction of the second cascade energies are smaller than \SI{8}{\gev} and as a result, for many events, only light from the first cascade was observed and therefore reconstructible. Overall, mostly single photo-electrons are detected in each DOM, while the total number of observed photo-electrons in an event peaks below 10.

\begin{figure}[h]
    \centering
    \includegraphics[width=\columnwidth]{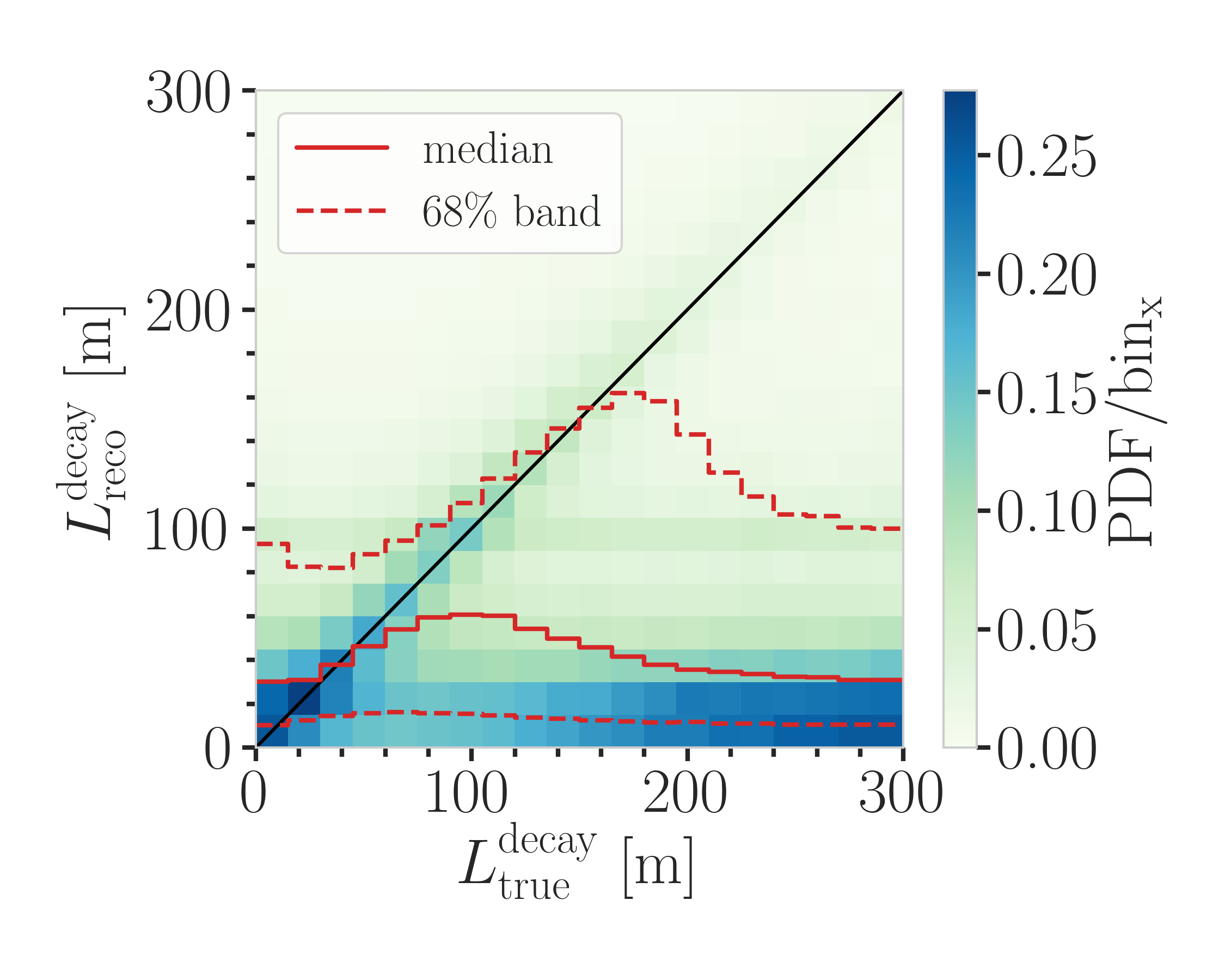}
    \caption{Reconstructed decay length versus true decay length, where the color scale is normalized per vertical slice and the median and \SI{68}{\percent} band are shown in red. For this development sample the decay lengths were sampled from a $L^{-1}$ distribution in the range from \SIrange{1}{1000}{\meter}, while the masses are sampled uniformly between \SIrange[range-phrase={~and~}]{0.1}{1.0}{\gev} and a mixing of $1.0$ is assumed.
    }
    \label{fig:taupede_reconstructed_decay_length}
\end{figure}

As a consequence of this, the decay length resolution (as shown in \cref{fig:taupede_reconstructed_decay_length}) only performs well for true decay lengths from \SIrange{20}{80}{\meter}, which can be seen by the median being almost unbiased on top of the diagonal. Nonetheless, the spread is still large at the order of $\pm \SI{40}{\meter}$ in the \SI{68}{\percent} band. Outside this region the reconstruction performs poorly, with the median not following the diagonal and above true decay lengths of \SI{80}{\meter}, in particular, more than \SI{60}{\percent} of the reconstructed decay lengths are below \SI{60}{\meter}. With increasing true decay lengths this fraction increases even more and it was found that these events were cases where only one cascade was observed with sufficient light to be reconstructed, while the other remained invisible, making it impossible to reconstruct the distance between the two.

In an attempt to identify HNL events over the SM background from neutrinos and $\mu$'s, the results from this reconstruction were fed into a multivariate machine learning method. The applied method was the Gradient boosting machine~\cite{BDT} from the \textsc{scikit-learn} package~\cite{sklearn} and further details of can be found in Ref.~\cite{Fischer2024First}. A minimum reconstructed length of \SI{40}{\meter} and reconstructed cascade energies above \SI{5}{\gev} was required for all events, while only a subset of signal events with true lengths between \SIrange{40}{150}{\meter} was used, significantly reducing their numbers. As a result of this, and given the overall limited reconstruction performance, the classification of the signal events over the large number of background events from SM atmospheric neutrino interactions was not successful.

%%%%%%%%%%%%%%%%%%%% Analysis Methodology %%%%%%%%%%%%%%%%%%%%

\section{\label{sec:analysis_methods}Analysis Methodology}

Theoretically, HNLs can produce a unique, low-energy double-cascade signature. Despite this fact, it was not possible to properly reconstruct them with the current selection and reconstruction tools available. Due to the lack of an effective reconstruction, we could then not classify the double-cascade events in a separate morphological category, to perform an analysis searching for occurrences of this unique signature. In the process it was however identified, that most of the events are observed as single cascades in our detector.
These events are observed on top of the SM background of atmospheric neutrino and $\mu$ events. For this reason, an analysis scheme was revised, that searches for this excess of mostly cascade-like events, without the need of explicitly identifying the events stemming from the BSM process.
In this alternative avenue to perform the search for HNLs, the standard reconstruction used for low-energy atmospheric neutrinos is applied to all events, including the signal MC, to perform a forward-folding-style analysis. In the case of a non-zero \ut4, this results in an excess of events from HNLs, observed on top of the expected events from SM interactions of atmospheric neutrinos and $\mu$'s which can be used to investigate the $m_4$-\ut4 space.

In this section, we first describe the final sample selection before introducing the binning scheme and the resulting background and signal shapes. After describing the fit method, the included systematic uncertainties are discussed at the end of the section.

\subsection{\label{sec:sample}Final Sample Selection}

All simulated and observed data are processed through the filtering chain described in \cref{sec:filter}, before being reconstructed with the algorithm explained in \cref{sec:flercnn}.
Using the processed simulation, an additional boosted decision tree (BDT) classifier is trained to separate neutrinos from atmospheric $\mu$'s as explained in Ref.~\cite{IceCube:2024xjj}.
The final selection is performed using the output probability of this BDT, in addition to a collection of spatial and hit-based variables.
The purpose of this selection is to increase the purity of the neutrino sample, to reject events with poor reconstruction quality, and to ensure good data to MC agreement.
The selection criteria are listed in \cref{tab:analysis_cuts} and their application leads to the final level sample.

\begin{table}[h]
    \footnotesize
        \begin{tabular}{ lcr }
        \hline\hline    
        \textbf{Variable} & \textbf{Threshold} & \textbf{Removed} \\
        \hline\hline    
        Hit DOMs & $\geq 7$ & \SI{1.05}{\percent} \\
        Radial distance & $< \SI{200}{\meter}$ & \SI{0.09}{\percent} \\
        Vertical position & $\SI{-495}{\meter} < z < \SI{-225}{\meter}$ & \SI{5.48}{\percent} \\
        Energy & $\SI{5}{\gev} < E < \SI{100}{\gev}$ & \SI{20.70}{\percent} \\    
        Cosine of zenith angle & $< 0.04$ & \SI{19.66}{\percent} \\
        Direct hits & $> 2.5$ & \SI{10.50}{\percent} \\
        Hits in top layers & $< 0.5$ & \SI{0.03}{\percent} \\
        Hits in outer layer & $< 7.5$ & \SI{0.001}{\percent} \\
        Muon classifier score & $\geq 0.8$ & \SI{23.90}{\percent} \\
        \hline
        \end{tabular}
    \caption{Selection criteria for the final analysis sample. Listed are the selection variables, the threshold or selection range, and the reduction of the background sample, when the selection of interest is applied after all other selections are already applied.}
    \label{tab:analysis_cuts}
\end{table}

\begin{figure}[h]
\includegraphics[width=\columnwidth]{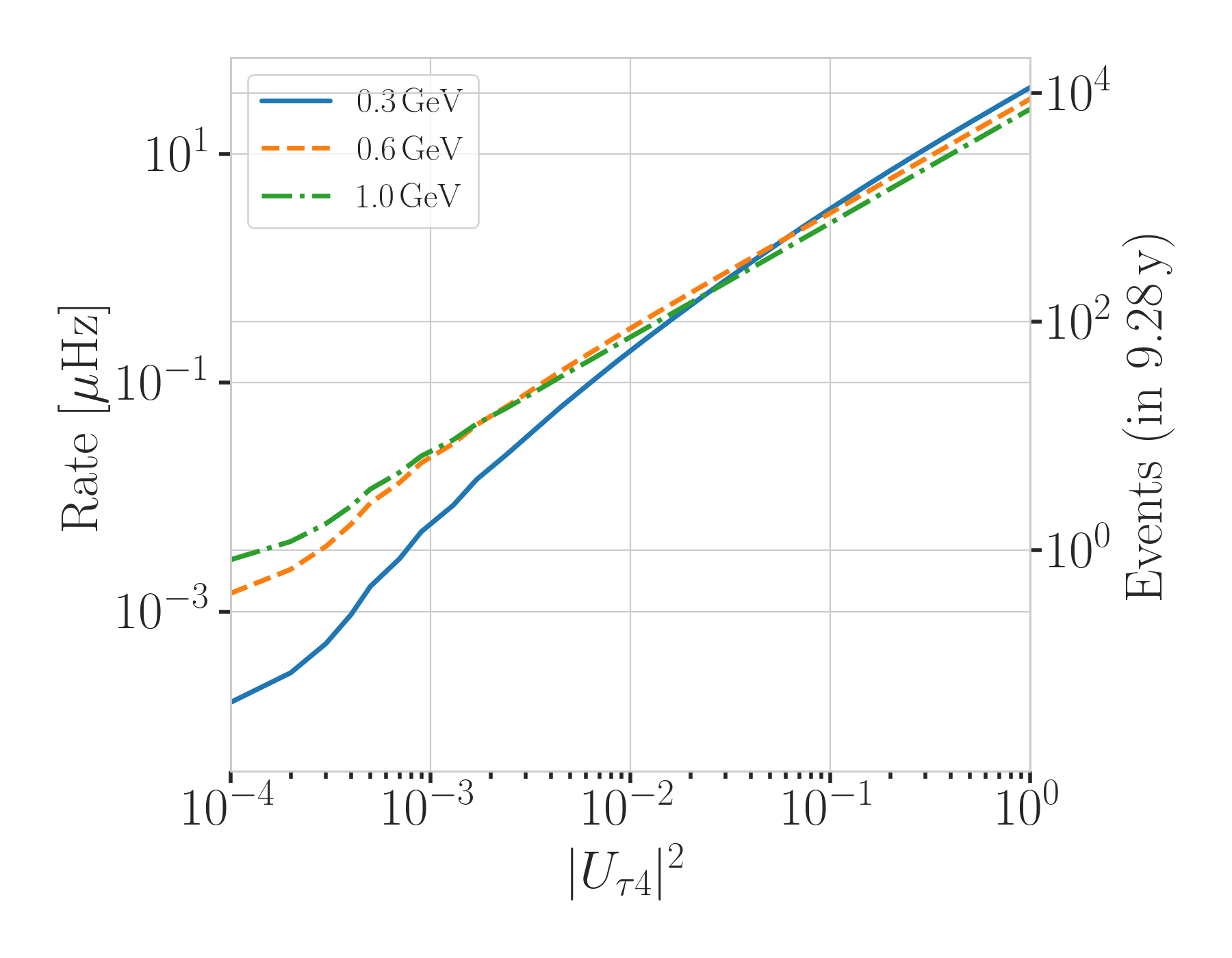}
\caption{HNL rates and event expectations for all three masses at final level of the event selection.}
\label{fig:signal_expecations}
\end{figure}

\cref{tab:background_final_level_expectation} shows the nominal rates and the nominal event expectation from SM backgrounds as well as the observed number of data events. The sample is neutrino-dominated, with the majority of events being from $\nu_\mu$-CC interactions. \cref{fig:signal_expecations} shows the signal expectation for the three simulated HNL masses, $m_4$ = {0.3, 0.6, 1.0}\,GeV, where both the rate and the expected number of events in \SI{9.28}{years} of data taking are shown with respect to the mixing. For reference, at a mixing of $\nomathut4=10^{-1}$, the HNL signal expectation is at a similar order as the number of expected atmospheric $\mu$'s in the final sample. The total rate scales roughly linearly with the mixing, where effects of the different visible energy fractions for the different mass samples affect the slope (shown in \cref{fig:hnl_branching_ratios_data}). Small mixing results in longer decay lengths, leading to non-linearities due to the limited Monte Carlo, which is only simulated up to a decay length of \SI{1000}{\meter}.

\begin{table}[h]
    \begin{tabular}{ ccrr }
    \hline\hline
    \textbf{Type} & \textbf{Rate [\si{\milli\hertz}]} & \textbf{Events (\SI{9.28}{years})} \\
    \hline\hline
    $\nu_\mu^\rm{CC}$   & 0.3531 & 103321 $\pm$ 113 \\
    $\nu_e^\rm{CC}$     & 0.1418 & 41490 $\pm$ 69 \\
    $\nu^\rm{NC}$       & 0.0666 & 19491 $\pm$ 47 \\
    $\nu_\tau^\rm{CC}$  & 0.0345 & 10094 $\pm$ 22 \\
    $\mu_\rm{atm}$      & 0.0032 & 936 $\pm$ 15 \\
    \hline
    total               & 0.5992 & 175332 $\pm$ 143 \\
    \hline
    data                & -      & 150257 $\pm$ 388 \\
    \hline
    \end{tabular}
\caption{Nominal rates and nominal event expectations of the SM background particle types and interactions assuming \SI{9.28}{years} of detector live time at final level of the analysis selection (pre-fit). The expected rate is calculated as the sum of the weights $\sum_i\omega_i$, with the stated uncertainty being $\sqrt{\sum_i\omega_i^2}$. Also shown is the number of observed data events in the same time period.}
\label{tab:background_final_level_expectation}
\end{table}

\subsection{\label{sec:signal_shape}Binning and Signal Shape}

\begin{figure*}[h]
    \includegraphics[width=\textwidth]{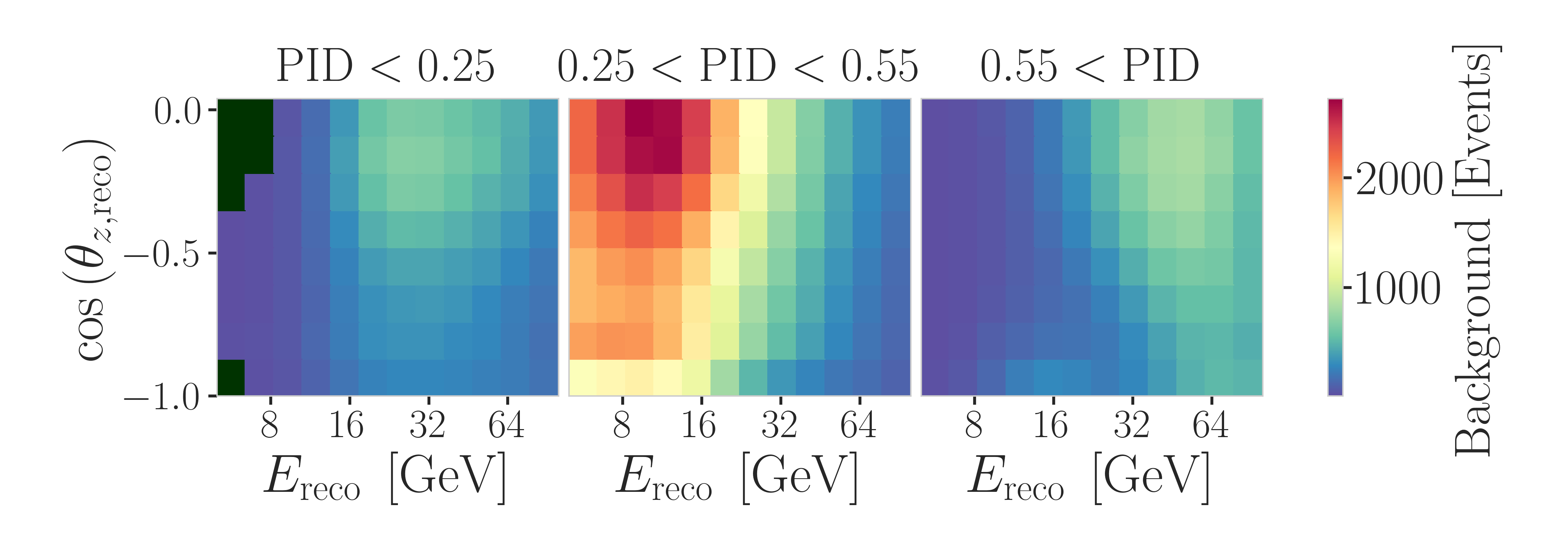}
% \begin{figure}[h]
    % \includegraphics[width=\columnwidth]{all_background.png}
    \caption{Nominal background expectation for \SI{9.28}{years} of live time as a function of reconstructed energy and direction for cascade-like, mixed, and track-like morphologies, from left to right respectively.}
    \label{fig:background_3d_hist}
\end{figure*}

\begin{figure*}[h]
    \includegraphics[width=\textwidth]{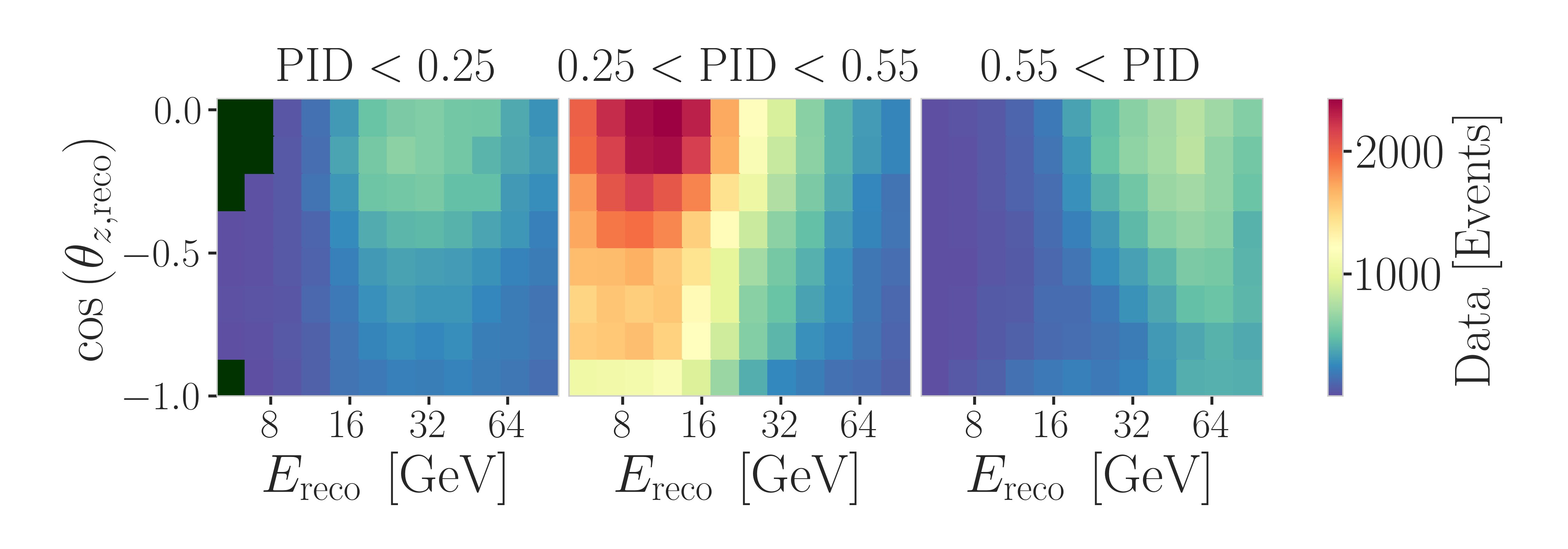}
% \begin{figure}[h]
    % \includegraphics[width=\columnwidth]{all_data.png}
    \caption{Observed data for \SI{9.28}{years} of live time, with the same format as \cref{fig:background_3d_hist}, where the background expectation is shown.}
    \label{fig:data_3d_hist}
    \end{figure*}

To enhance the sensitivity to the HNLs' specific event signature, we further divide the events based on three reconstructed variables.
There are twelve bins in reconstructed energy, eight bins in cosine of the reconstructed zenith angle, and three bins in PID, which correspond to track-like ($\mathrm{PID} > 0.55$), mixed ($0.25 < \mathrm{PID} < 0.55$), and cascade-like ($\mathrm{PID} < 0.25$) morphologies.
The binning scheme is summarized in \cref{tab:analysis_binning} and is identical to the binning used in Ref.~\cite{IceCube:2024xjj}.
The nominal background expectation in these bins is shown in  \cref{fig:background_3d_hist}.
During the binning scheme development we found that some lower energy bins in the cascade-like PID bin have small expected numbers of events, and can not be adequately described by Gaussian statistics.
These bins are therefore removed from the analysis and illustrated in dark green in \cref{fig:background_3d_hist} and all similar figures that follow. The data observed in \SI{9.28}{years} of live time, binned in the same scheme, is shown in \cref{fig:data_3d_hist}

% \end{figure}

\begin{figure*}[h]
    \includegraphics[width=\textwidth]{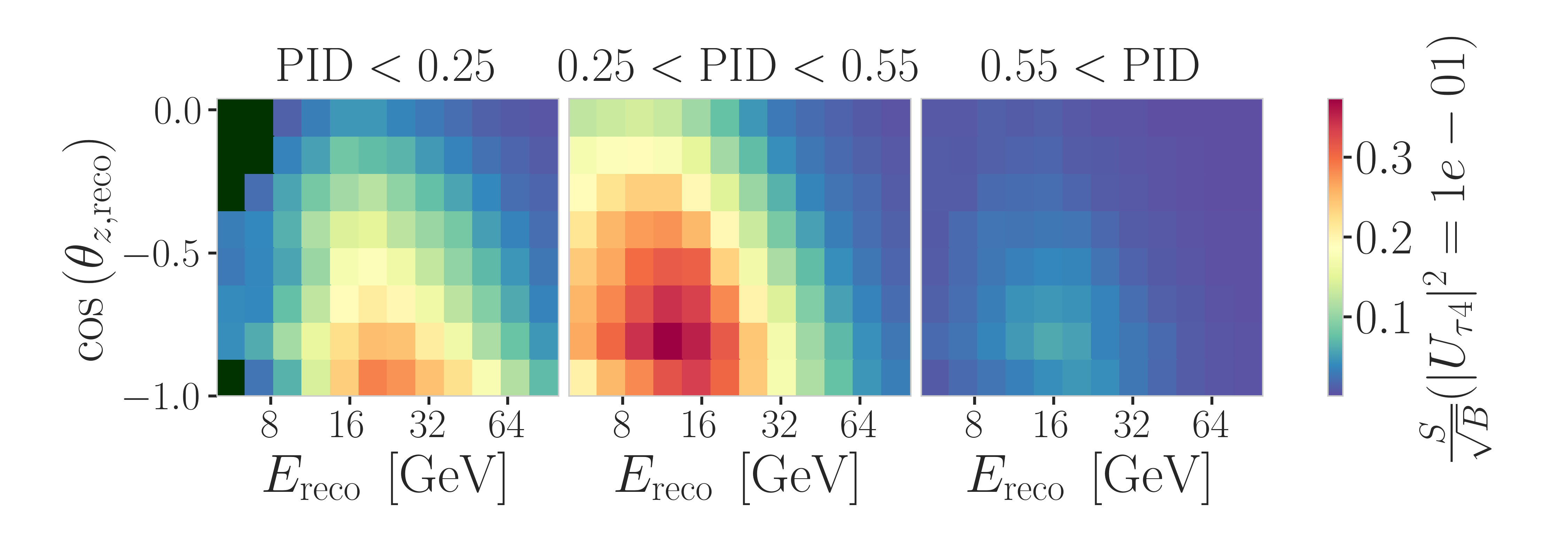}
% \begin{figure}[h]
    % \includegraphics[width=\columnwidth]{labeled_s_to_sqrt_b_1_0_GeV_combined_U_tau4_sq_0.1000_total.png}
    \caption{Signal events divided by the square root of the nominal background expectation in \SI{9.28}{years}. This assumes a \SI{1.0}{\gev} HNL mass and a mixing of $\nomathut4=10^{-1}$.}
    \label{fig:s_to_sqrt_b_1.0_GeV_0.1_mixing}
\end{figure*}
% \end{figure}

\begin{table}[h]
    \small
    \begin{tabular}{ llll }
    \hline\hline    
    \textbf{Variable} & \textbf{$N_\rm{bins}$} & \textbf{Edges} & \textbf{Spacing} \\     
    \hline\hline    
    $P_\nu$ & 3 & [0.00, 0.25, 0.55, 1.00] & linear \\
    $E$ & 12 & [5.00, 100.00] & logarithmic \\
    $\cos(\theta)$ & 8 & [-1.00, 0.04] & linear \\    
    \hline
\end{tabular}
\caption{Three-dimensional binning used in the analysis. All variables are from the CNN reconstruction explained in \cref{sec:filter_and_reco}.}
\label{tab:analysis_binning}
\end{table}

\cref{fig:s_to_sqrt_b_1.0_GeV_0.1_mixing} shows the expected signal shape in the analysis binning assuming an HNL mass of \SI{1.0}{\GeV} and a mixing of $\nomathut4=10^{-1}$. As expected, the signal shows up mostly in the cascade and mixed bins, since for many of the events only a single-cascade deposits light in the detector. A small fraction of the events falls into the track-like bin, which can be explained by them having a more elongated light deposition pattern, due to the two cascades, and therefore being classified as track-like by the CNN. It can also be seen that the signal is concentrated in the lower energy bins, at and below the energy at which the $\nu_\tau$ flux peaks. As discussed in detail in \cref{sec:generator}, a fraction of the energy goes into invisible particles and does not deposit any light, therefore lowering the reconstructed energy of the events compared to the energy of the incoming $\nu_\tau$. The distribution of the signal events into the cascade-like, mixed, and track-like bins is dependent on both the mixing value as well as the mass, because both modify the proper lifetime of the HNL and with it the decay length in the detector. For very short and very large decay lengths, the events look more cascade-like, while for intermediate lengths, the events can be more elongated, leading to a classification as more track-like. Distributions for all masses and a two mixings of $\nomathut4=10^{-1}$ and $\nomathut4=10^{-3}$ can be found in \cref{app:signal_shapes} in addition to ratio comparisons of the different mass samples, highlighting their kinematic differences.

% \end{figure}

\subsection{\label{sec:analysis_setup}Fit Method}

The search for HNLs is performed through a binned maximum likelihood estimation. The parameters of interest in this study are the mass of the additional heavy state, $m_4$, and the mixing matrix element \ut4. We perform independent searches for $m_4$ = {0.3, 0.6, 1.0}\,GeV, where the mixing \ut4 is varied continuously in each fit.

To compare the bin-wise expectations to the data, a modified $\chi^2$ is used as the fit metric. It is defined as
\begin{equation}
    \footnotesize
    \chi^2_{\mathrm{mod}} =
    \sum_{i \in \mathrm{bins}}^{}\frac{(N^{\rm{exp}}_i - N^{\mathrm{obs}}_i)^2}
    {N^{\rm{exp}}_i + (\sigma^{\mathrm{\nu}}_i)^2 + (\sigma^{\mathrm{\mu}}_i)^2 + (\sigma^{\mathrm{HNL}}_i)^2}
    + \sum_{j \in \mathrm{syst}}^{}\frac{(s_j - \hat{s_j})^2}{\sigma^2_{s_j}}
    \;,
    \label{eq:mod-chi2-hnl}
\end{equation}
where $N^{\rm{exp}}_i = N^{\mathrm{\nu}}_i + N^{\mathrm{\mu}}_i + N^{\mathrm{HNL}}_i$ is the total event expectation in bin $i$, with $N^{\mathrm{\nu}}_i$, $N^{\mathrm{\mu}}_i$, and $N^{\mathrm{HNL}}_i$ being the expected number of events from neutrinos, atmospheric $\mu$'s, and HNLs, and $N^{\mathrm{obs}}_i$ is the observed number of data events. Summing the weights of all events of a particular type gives the expected number of events as $N^{\mathrm{type}}_i = \sum_i^\rm{type}\omega_i$, with the statistical uncertainty being $(\sigma^{\mathrm{type}}_i)^2 = \sum_i^\rm{type}\omega_i^2$, which is included in the fit metric to account for the limited MC statistics. A Gaussian penalty is added for each systematic parameter, $s_j$, with a known uncertainty, forming the additional term in \cref{eq:mod-chi2-hnl}. The nominal value of the parameter is given by $\hat{s_j}$, with a prior width of $\sigma_{s_j}$. The systematic uncertainties that are included as nuisance parameters in the fit are discussed in detail in \cref{sec:uncertainties}. Both their nominal values and priors (if applicable) are listed in the first column of \cref{tab:best_fit_parameters}.

To conduct the fit, we use the \textsc{PISA}~\cite{pisa_paper, pisa_software} software framework.
It was explicitly designed to perform analyses of small signals in high-statistics experiments and is employed to generate the expected event distributions from several MC samples, compare them to the observed data, and perform the minimization.
Before performing a fit to the real detector data, the analysis method is validated using Asimov pseudo-data generated from MC.
The minimization is done using the \textit{MIGRAD} routine of \textsc{iminuit}~\cite{iminuit_v2.17.0}, in  a stage-wise manner.
First, a fast fit with coarse minimizer settings is used to locate an estimate of the best-fit point (BFP), and second, a fine fit with more precise minimizer settings is performed starting from the BFP. The minimizer settings for both steps are shown in \cref{tab:minimization_settings}.

\begin{table}[h]
    % \small
        \begin{tabular}{ lcrc }
        \hline\hline
        \textbf{Fit} & \textbf{Error} & \textbf{Precision} & \textbf{Tolerance} \\
        \hline\hline
        Coarse & 1e-1 & 1e-8 & 1e-1 \\
        Fine & 1e-5 & 1e-14 & 1e-5 \\
        \hline
        \end{tabular}
    \caption{Migrad settings for the two stages in the minimization routine, showing the step size for the numerical gradient estimation, the precision of the LLH calculation, and the tolerance.}
    \label{tab:minimization_settings}
\end{table}

The minimization routine is validated by checking the convergence for a broad range of physics parameters through fits to pseudo-data generated without statistical fluctuations.
The fits always recover the true physics parameter within the minimizer tolerance.

\subsection{\label{sec:uncertainties} Systematic Uncertainties}

The included nuisance parameters stem from several sources of systematic uncertainties related to both event generation and detector response. Following the work presented in Ref.~\cite{IceCubeCollaboration:2023wtb}, the physical models are used to implement the uncertainties on event distributions, resulting from varying these systematic parameters. For the cases where no analytical description is available, effective parameters, calculated based on Monte Carlos simulations, are used. We list here only the systematic uncertainties that were found to be relevant for this analysis, while all other parameters are kept fixed at their nominal values~\cite{Fischer_2023}.

All uncertainties that scale the total neutrino expectation are included in a single scaling parameter $N_{\nu}$, because the analysis is primarily sensitive to the relative distribution of events, while the total flux of neutrinos is not relevant for the measurement itself.
Three parameters, $h_{\pi^+}$, $i_{\pi^+}$, and $y_{K^+}$, are used to account for the uncertainties in the pion and kaon production in the atmosphere, which are the dominant sources of neutrino production at the energies considered in this work.
They are a selection of the effective parameters introduced in Ref.~\cite{Barr:2006it} and are constrained by Gaussian priors.
The remaining parameters in this parameterization were found to have a negligible impact on the analysis and are fixed to their nominal values.
The correction factor $\Delta \gamma_\nu$ accounts for the uncertainty in the cosmic ray flux through a power-law correction, choosing a prior of 0.1, based on recent measurements from Ref.~\cite{PhysRevD.95.023012}.
No correction yields the prediction from the baseline model taken from Ref.~\cite{PhysRevD.92.023004_Honda_Flux}.

Cross-section uncertainties are divided into low- and high-energy components, since different processes contribute at different regimes.
Two parameters $M_\rm{A,QE}$, and $M_\rm{A,RES}$ account for the uncertainties in form factors of charged-current quasi-elastic (CCQE) and charged-current resonant (CCRES) interactions at low energies.
The third parameter, $\rm{DIS}$, accounts for the disagreement between the \textsc{GENIE} and CSMS~\cite{csms} models for the deep-inelastic scattering (DIS) cross-sections at higher energies.

The uncertainties that cannot be described by analytical models are the detector systematic uncertainties, which have a strong impact on the analysis.
A DOM light detection efficiency scale, $\epsilon_{\rm{DOM}}$, is included to account for the overall uncertainty in the photon detection of the modules.
This parameter is constrained by a Gaussian prior with a width of $0.1$, based on the studies from Refs.~\cite{JFeintzeig_phd, domeff_nick}.
Two additional parameters, $\rm{hole \, ice} \, p_0$, and $\rm{hole \, ice} \, p_1$, related to relative efficiencies based on the photons' incidence angles, are used to account for uncertainties due to the local optical properties of the re-frozen ice columns called hole ice~\cite{Eller:2023ko}.
Global ice property effects are accounted for by the $\rm{ice \, absorption}$, and $\rm{ice \, scattering}$ parameters, which scale the absorption and scattering lengths of the ice, respectively.
Additionally, to account for recent advancements in the ice model descriptions, that incorporate effects of the birefringent polycriystalline microstructure of the ice~\cite{bfr_ice_tc-18-75-2024}, a parameter, $N_\rm{bfr}$, is included to account for this birefringence (BFR) effect, which interpolates between the baseline ice model used in this analysis and the new model~\cite{IceCube:2024xjj}.

Since no analytical descriptions exist to describe the effect of the detector's systematic uncertainties, effective re-weighting parameters are derived from MC simulations through a novel method applying a likelihood-free inference technique.
Based on the MC samples that were generated with discrete variations of the detector systematic parameters, re-weighting factors are found for every single event in the nominal MC sample.
The re-weighting factors quantify the change in the relative importance of an event for the specific variation of detector systematic parameters. The method was initially introduced in Ref.~\cite{Fischer_2023} and first used in a light sterile neutrino search described in Ref.~\cite{Abbasi:2024ktc}.

The oscillation parameters $\theta_{23}$ and $\Delta m^{2}_{31}$, which govern the atmospheric neutrino oscillations, are also included in the fit with nominal values of \SI{47.5}{\degree} and \SI{2.48e-3}{\electronvolt^2}, respectively, chosen based on the results from the latest IceCube neutrino oscillation analysis~\cite{IceCube:2024xjj}.
They are relevant for this search because they govern the shape and the strength of the $\nu_\tau$ flux, stemming from the oscillation of $\nu_\mu$ to $\nu_\tau$.
No prior is applied to these parameters, allowing them to fit freely.
The mass ordering is fixed to normal ordering, since there is no dependency of this analysis on the ordering.

%%%%%%%%%%%%%%%%%%%% Results %%%%%%%%%%%%%%%%%%%%

\section{\label{sec:results}Results}

After the analysis procedure has been validated using Monte Carlo simulation, we perform three fits to the data, by minimizing \cref{eq:mod-chi2-hnl}, one for each value of $m_{4}$ being tested.
The best-fit mixing parameters for each HNL mass fit are found at
\begin{align*}
    \nomathut4(\SI{0.3}{\gev}) &= 0.003 \;, \\
    \nomathut4(\SI{0.6}{\gev}) &= 0.080 \;, \rm{and} \\
    \nomathut4(\SI{1.0}{\gev}) &= 0.106 \;.
\end{align*}
We provide the $p$-value to reject the NH for each mass in \cref{tab:best_fit_parameters_and_confidence_levels}, along with the upper limits on \ut4. 
\begin{figure}[h]
    \includegraphics[width=0.95\columnwidth, trim=0 0.5cm 0 0.5cm, clip]{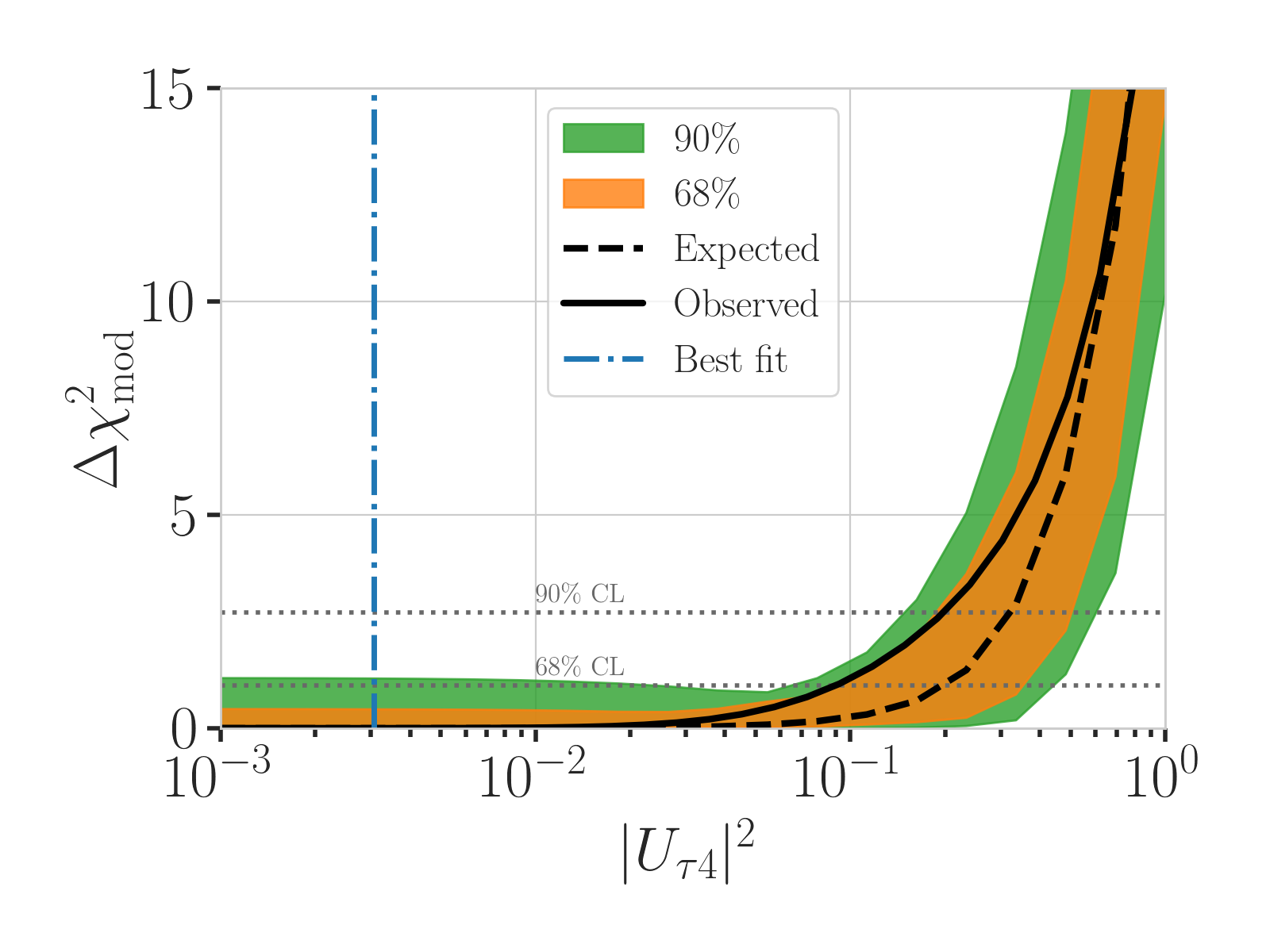}
    \includegraphics[width=0.95\columnwidth, trim=0 0.5cm 0 0.5cm, clip]{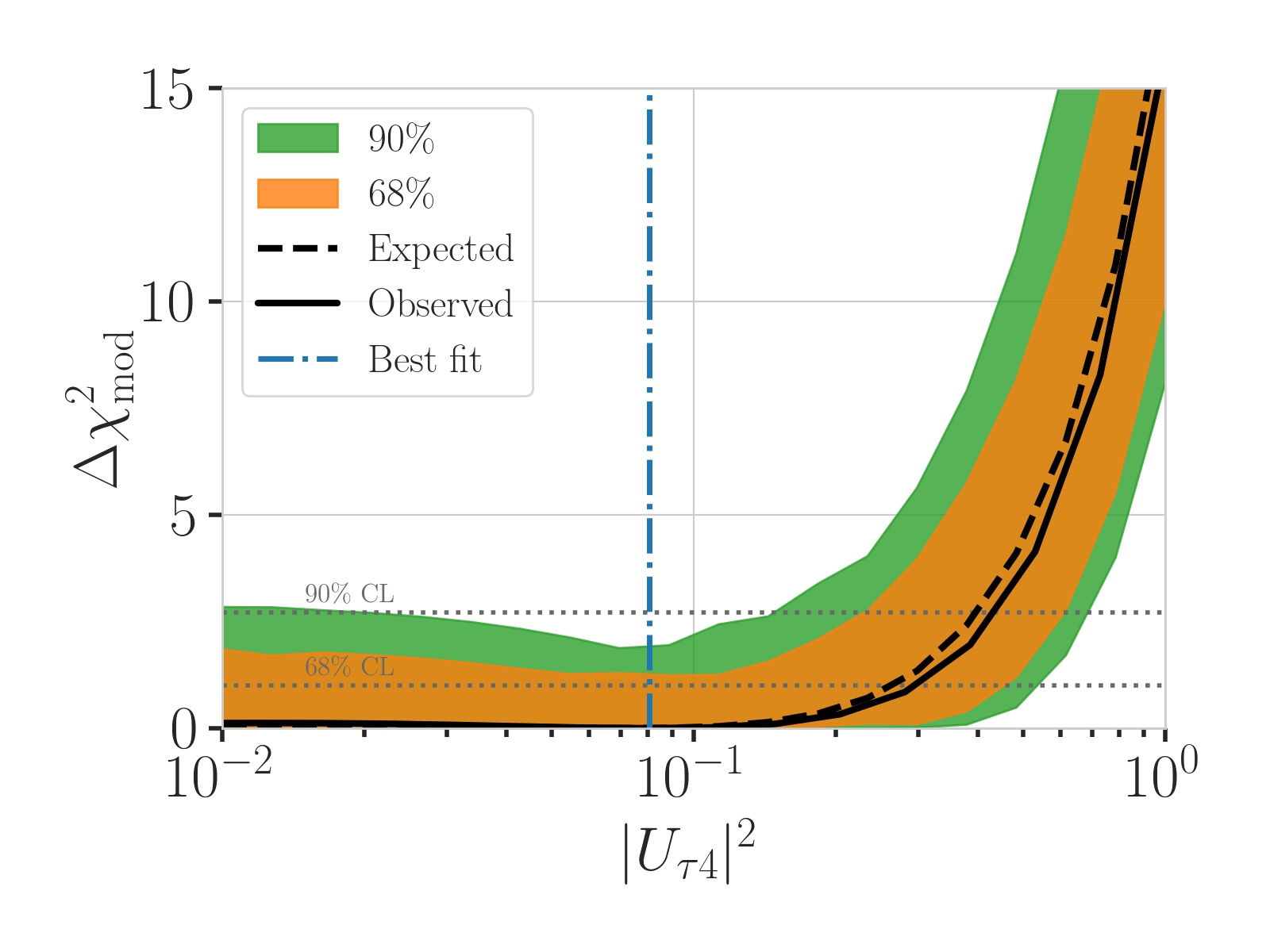}
    \includegraphics[width=0.95\columnwidth, trim=0 0.5cm 0 0.5cm, clip]{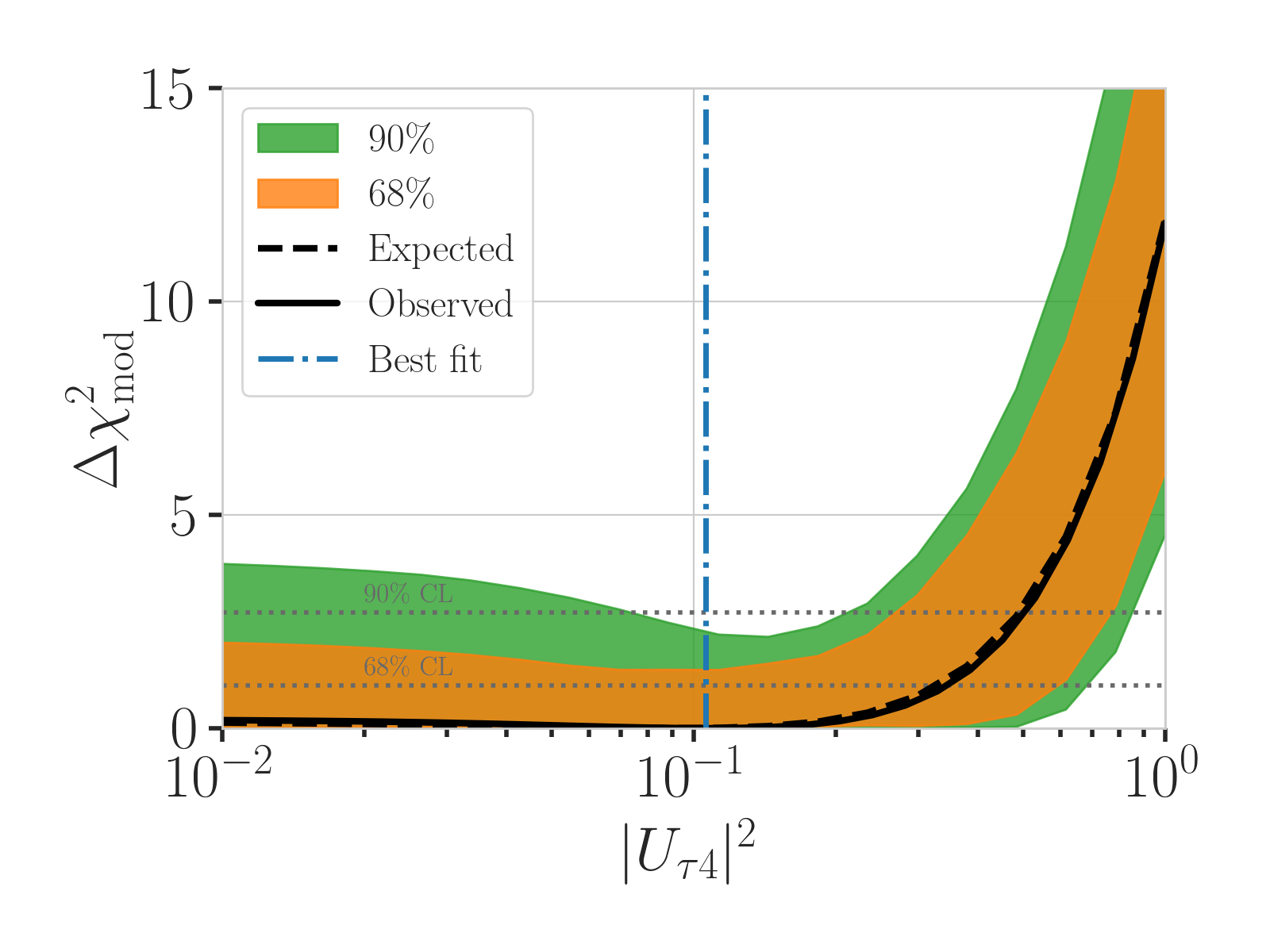}
% \begin{figure*}[h]
    % \includegraphics[width=0.6\textwidth, trim=0 0.5cm 0 0.5cm, clip]{brazil_band_with_asimov_0.3_GeV_updated_with_bfp_with_1sigma.png}
    % \includegraphics[width=0.6\textwidth, trim=0 0.5cm 0 0.5cm, clip]{brazil_band_with_asimov_0.6_GeV_updated_with_bfp_with_1sigma.png}
    % \includegraphics[width=0.6\textwidth, trim=0 0.5cm 0 0.5cm, clip]{brazil_band_with_asimov_1.0_GeV_updated_with_bfp_with_1sigma.png}
    \caption[Best fit point TS profiles]{Best fit point TS profiles as a function of \ut4 for the \SI{0.3}{\gev} (top), \SI{0.6}{\gev} (middle), and \SI{1.0}{\gev} (bottom) mass samples. Shown are the observed profiles, the Asimov expectation at the best fit point, and the \SI{68}{\percent} and \SI{90}{\percent} bands, based on 100 pseudo-data trials. Also indicated are the \SI{68}{\percent} and the \SI{90}{\percent} confidence levels assuming Wilks' theorem (horizontal dotted lines).}
    \label{fig:brazil_bands}
% \end{figure*}
\end{figure}
These results assume that Wilks' theorem holds~\cite{wilks_theorem}, which was checked at the 68\% level and found to hold well.
They are all compatible with the null hypothesis (NH) of no mixing, i.e., $\nomathut4=0.0$.
\begin{table}[h]
    \begin{tabular}{ ccccc }
        \hline\hline
        \textbf{HNL mass} & \textbf{\ut4} & \textbf{68\% CL} & \textbf{90\% CL} & \textbf{NH $p$-value} \\    
        \hline\hline
        \SI{0.3}{\gev} & 0.003 & 0.09 & 0.19 & \SI{0.97}{} \\
        \SI{0.6}{\gev} & 0.080 & 0.21 & 0.36 & \SI{0.79}{} \\
        \SI{1.0}{\gev} & 0.106 & 0.24 & 0.40 & \SI{0.63}{} \\
        \hline
    \end{tabular}
    \caption{Best-fit mixing values and the corresponding upper limits at the \SI{68}{\percent} and the \SI{90}{\percent} confidence level, as well as the $p$-value to reject the null hypothesis, estimated using Wilks' theorem.}
    \label{tab:best_fit_parameters_and_confidence_levels}
\end{table}
The observed test statistic (TS) profiles are shown in \cref{fig:brazil_bands}. The TS is calculated as the $\Delta\chi^2_{\mathrm{mod}}$ between a free fit and a fit where the mixing is fixed to a given value.
The expected Asimov profile and its fluctuations based on 100 pseudo-data trials are also shown, where the pseudo-data was generated assuming the best-fit physics and nuisance parameters, fluctuating the bin counts according to Poisson statistics and the MC statistical uncertainty.
The observed profiles lie within the \SI{68}{\percent} band, confirming their compatibility with statistical fluctuations of the data.

\begin{figure}[h]
    \includegraphics[width=0.9\columnwidth, trim=0 0.5cm 0 1.5cm, clip]{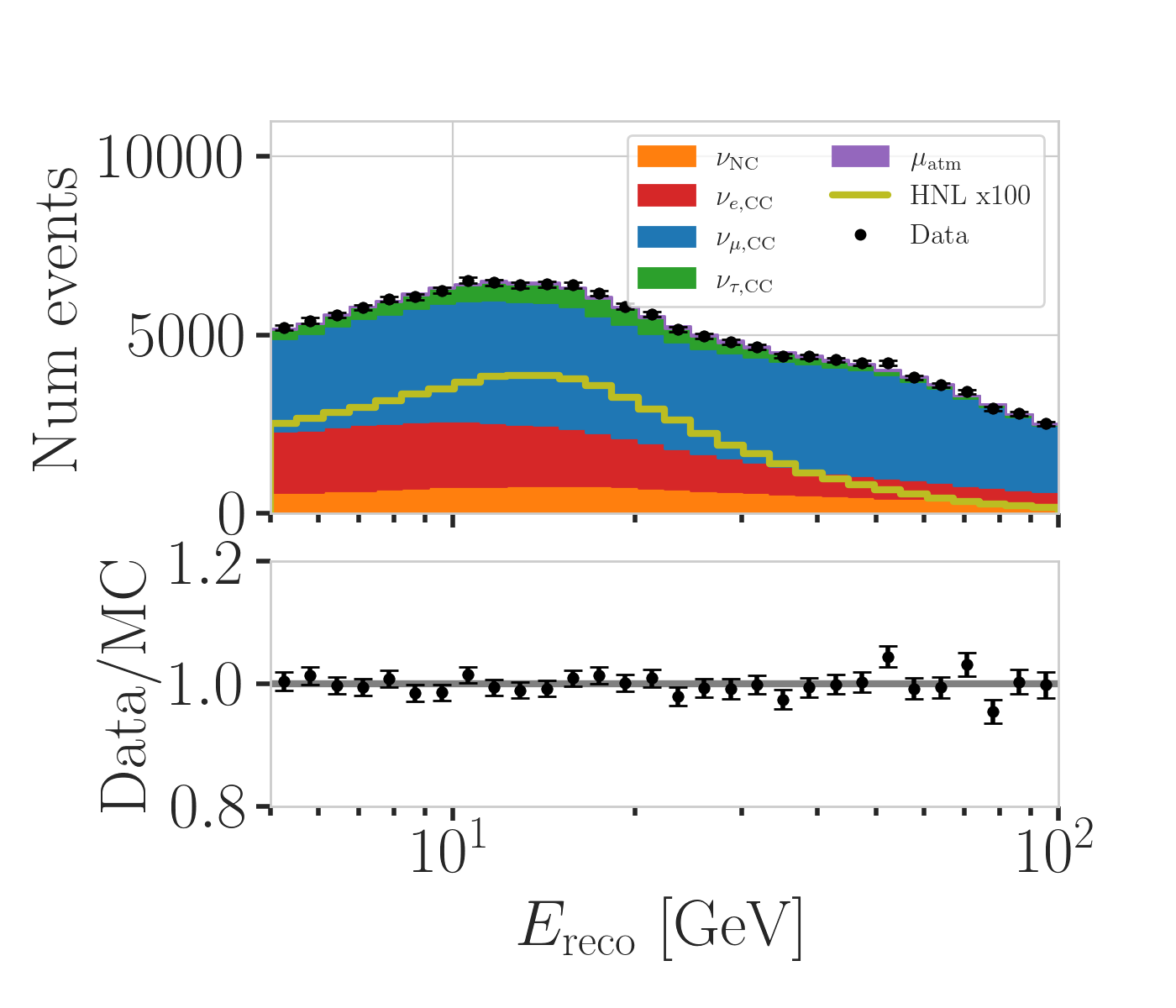}
    \includegraphics[width=0.9\columnwidth, trim=0 0.5cm 0 1.5cm, clip]{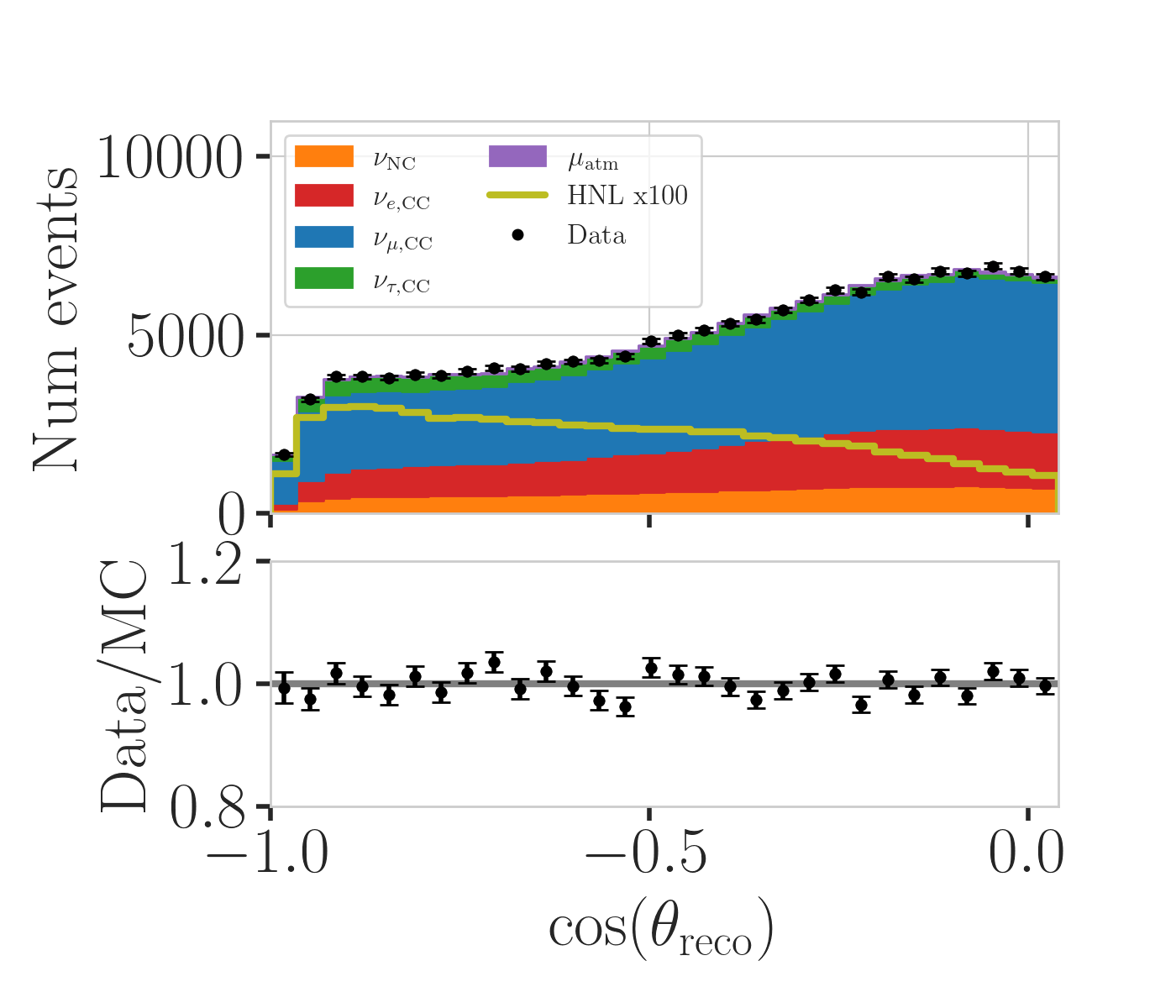}
    \includegraphics[width=0.9\columnwidth, trim=0 0.5cm 0 1.5cm, clip]{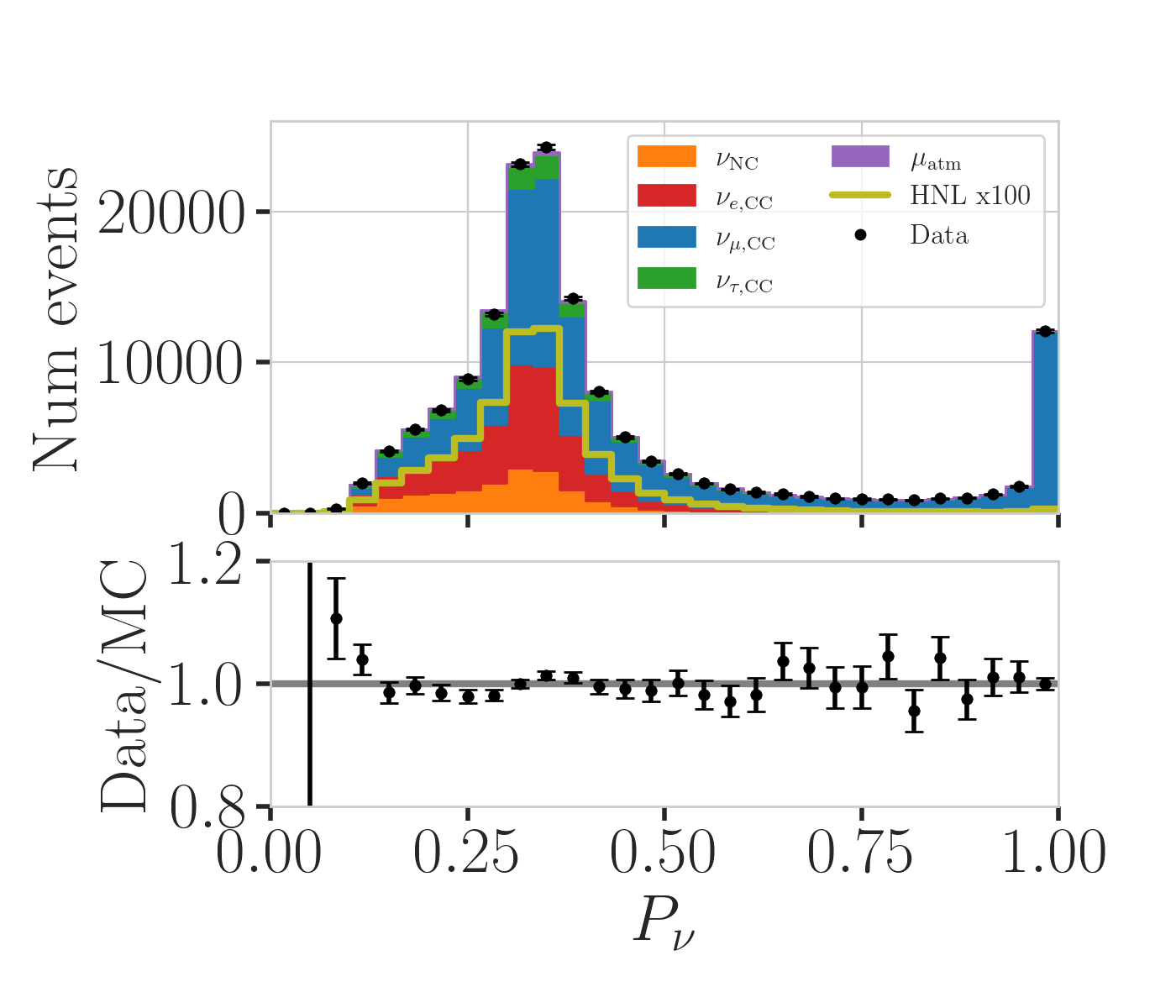}
    \caption{Distributions of reconstructed energy, cosine of the reconstructed zenith angle, and PID of the data (black points) compared to the best fit MC shown as stacked histograms for different particle interaction types. The HNL events, which assume an HNL mass of 0.6 GeV, are included as part of the $\nu_\mathrm{NC}$ events in the stacked histogram, and also shown separately multiplied by a factor of 100 in gold.}
    \label{fig:1_d_data_mc_bfp_0.6_GeV_combined}
\end{figure}

The one-dimensional distributions of the reconstructed energy, PID, and cosine of the zenith angle for data and MC are shown in \cref{fig:1_d_data_mc_bfp_0.6_GeV_combined}.
Here, we show the results of the fit assuming an HNL mass of $\SI{0.6}\GeV$, but the distributions from the 0.3 and $\SI{1.0}\GeV$ masses are similar, because no significant signal was found and the BFP background distributions agree within the statistical uncertainty.
The bottom panel highlights the excellent agreement between data and simulation for all three variables, with the ratio being well centered around unity and the spread being small at the \si{\percent} level.
The HNL signal is shown as part of the $\nu_\mathrm{NC}$ events in the stacked histogram, but also separately multiplied by a factor of 100, to show its shape.
As expected by the nature of the production from $\nu_\tau$ oscillations, the signal is mostly concentrated at low energies and in the cascade-like PID region, while its direction is mostly upward-going.

\begin{table}[h]
    \begin{tabular}{ ll lll }
    \hline\hline
    \textbf{Parameter} & \textbf{Nominal} & \multicolumn{3}{c}{\textbf{Best-Fit / \textit{Pull} [\si{\sigma}]}} \\ 
    m$_{4}$ (fixed) & - & \textbf{\SI{0.3}{\gev}} & \textbf{\SI{0.6}{\gev}} &  \textbf{\SI{1.0}{\gev}} \\ 
    \hline\hline
    $|U_{\tau 4}|^2$                                 & -                    & 0.003               & 0.080               & 0.106               \\
    GOF p-value                                      & -                    & \SI{28.3}{\percent} & \SI{28.7}{\percent} & \SI{26.0}{\percent} \\
    \hline

    $\theta_{23} [\si{\degree}]$                     &    47.5              &          48.1       &          47.9       &          48.0       \\ 
    $\Delta m^{2}_{32} [\SI{e-3}{\electronvolt^2}] $ &    2.401             &              2.380  &              2.380  &              2.381  \\ 
    \hline
    $N_{\nu}$                                        &    1.0               &           0.889     &           0.889     &           0.890     \\ 

    $\Delta \gamma_\nu$                              &    0.0$\pm$0.10      &          \textit{-0.079}    &          \textit{-0.067}    &          \textit{-0.066}      \\ 
    $h_{\pi^+}$                                      &    0.0$\pm$0.15      &          \textit{-0.98}     &          \textit{-0.99}     &          \textit{-0.99}      \\ 
    $i_{\pi^+}$                                      &    0.0$\pm$0.61      &           \textit{0.78}     &          \textit{0.84}      &           \textit{0.86}      \\ 
    $y_{K^+}$                                        &    0.0$\pm$0.30      &           \textit{0.25}     &          \textit{0.21}      &           \textit{0.19}      \\ 
    \hline
    $\rm{DIS}$                                       &    0.0$\pm$1.0       &          \textit{-0.25}     &          \textit{-0.22}     &          \textit{-0.22}      \\ 
    $M_\rm{A,QE}$                                    &    0.0$\pm$1.0       &          \textit{-0.17}     &          \textit{-0.13}     &          \textit{-0.12}      \\ 
    $M_\rm{A,res}$                                   &    0.0$\pm$1.0       &          \textit{-0.13}     &          \textit{-0.08}     &          \textit{-0.07}      \\ 
    \hline
    $\epsilon_{\rm{DOM}}$                            &    1.0$\pm$0.1       &           \textit{0.22}     &           \textit{0.18}     &           \textit{0.17}      \\ 

    $\rm{hole \, ice} \; p_0$                        &    0.102             &          -0.161     &          -0.161     &          -0.160     \\ 
    $\rm{hole \, ice} \; p_1$                        &   -0.049             &          -0.074     &          -0.076     &          -0.076     \\ 
    $\rm{ice \, absorption}$                         &    1.00              &           0.943     &           0.942     &           0.942     \\ 
    $\rm{ice \, scattering}$                         &    1.05              &           0.986     &           0.989     &           0.989     \\ 
    $N_\rm{bfr}$                                     &    0.0               &           0.747     &           0.740     &           0.736     \\

    \hline
    \end{tabular}
\caption{Best fit nuisance parameters for the three mass samples. Shown are the nominal values, the best-fit values for parameters without a prior, or the pull in units of $\sigma$ for parameters that have a (Gaussian) prior.}
\label{tab:best_fit_parameters}
\end{table}

\begin{figure}[h]
    \includegraphics[width=\columnwidth]{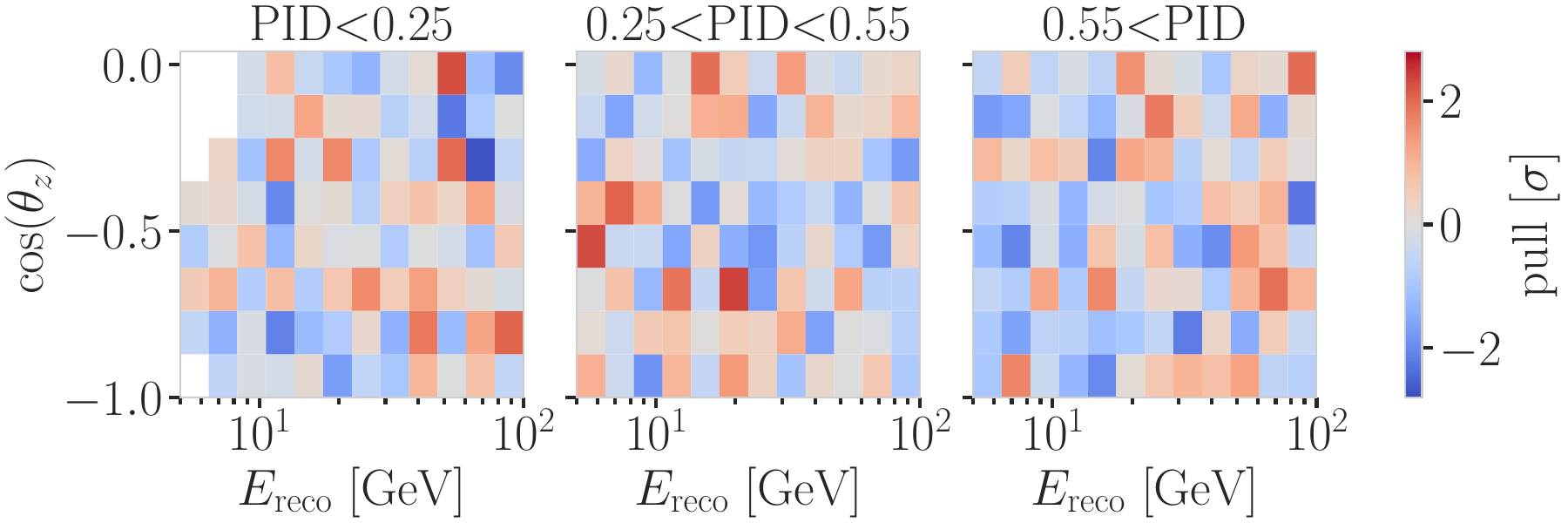}
    	\caption{Three-dimensional bin-wise pulls between data and simulation at the best fit point of the \SI{0.6}{\gev} mass sample fit.}
    \label{fig:3_d_bfp_pull_0.6_GeV}
\end{figure}

Additionally, all three best-fit $\chi^2_{\mathrm{mod}}$ values are found to be compatible with the expectation derived from an ensemble of pseudo-data trials for the corresponding HNL mass.
The ensemble is generated by injecting the BFP parameters and fluctuating the bin counts accounting for both MC statistical uncertainty and Poisson fluctuations in data, before fitting it with the same analysis chain as for the real data.
The goodness-of-fit (GOF) is estimated by comparing the final fit metric value to the distribution of values expected from the ensemble, resulting in a p-value, quantifying the probability of finding this result. 
The GOF p-values are listed in \cref{tab:best_fit_parameters}. For further bias checks, bin-wise pulls between data and the best-fit MC are investigated, illustrated in \cref{fig:3_d_bfp_pull_0.6_GeV} for the $m_{4}=\SI{0.6}{\gev}$ fit.
The pulls are centered around zero and follow a Normal distribution, further demonstrating good agreement between data and MC.
Similar results are obtained for the other masses.

The best-fit nuisance parameters are listed in \cref{tab:best_fit_parameters}, showing that all fits prefer similar values, and parameters with a prior are within their $\SI{1}{\sigma}$ range.
The atmospheric neutrino oscillation parameters vary by $<\SI{1}{\percent}$ for $\Delta m^2_{32}$ and by $\sim\SI{4}{\percent}$ for $\theta_{23}$, and are consistent with the results recently reported in~\cite{IceCube:2024xjj}.
This agreement is expected, given that the same sample is used in both analyses, with only small differences in the treatment of systematic uncertainties.

%%%%%%%%%%%%%%%%%%%% Conclusions and Future Prospects %%%%%%%%%%%%%%%%%%%%

\section{\label{sec:conclusion}Conclusions}

In this article, we present the first search of GeV-scale heavy neutral leptons with a neutrino telescope, using ten years of IceCube data.
We have introduced a new, publicly available event generator for HNL simulation in neutrino telescopes called \textsc{LI-HNL}, which was built on the \textsc{LeptonInjector} event generator.
We discuss the challenges and development of a dedicated HNL reconstruction, which seeks to find a double-cascade morphology that would be a smoking-gun signature of this particle if observed in sub-\SI{100}{\gev} neutrinos.
Unfortunately, given the low light yield and the sparse detector instrumentation, with the current tools we are not able to successfully identify low-energy double-cascade events against the background from SM atmospheric neutrino interactions.
The fact that we presently cannot single out double-cascade events from the background implies that the sensitive region obtained in Ref.~\cite{Coloma:2017ppo}, by assuming an order-one events detected, is significantly different from the sensitive region in this paper.
Given these findings, we instead perform an analysis looking for excess events in the energy and angular distributions using standard low-energy morphological categories in IceCube.

Our analysis focused on three masses: \SI{0.3}{\gev}, \SI{0.6}{\gev}, and \SI{1.0}{\gev}. 
The best-fit point at each mass was found to be consistent with the null hypothesis of no HNL mixing.
We place upper limits on the $\tau$-mixing of HNLs at $\nomathut4 < 0.19\;(m_4 = \SI{0.3}{\gev})$, $\nomathut4 < 0.36\;(m_4 = \SI{0.6}{\gev})$, and $\nomathut4 < 0.40\;(m_4 = \SI{1.0}{\gev})$ at the \SI{90}{\percent} confidence level.
Despite these constraints being several orders of magnitude below the current leading limits on \ut4 from reinterpretations of the CHARM and the BEBC results, which are at the order of $10^{-3}$ to $10^{-6}$ in the \SIrange{0.1}{2}{\gev} range~\cite{Orloff:2002de, Boiarska:2021yho, Barouki:2022bkt}, this initial result serves as a successful \textit{proof-of-concept} for HNL searches both in IceCube in particular and using atmospheric neutrinos in general.
Lastly, the future low-energy extension IceCube Upgrade~\cite{Ishihara:2019aao} will significantly improve the sensitivity to low-energy events through reduced spacing and segmented directionality of the optical modules.
This enhances light detection and should yield a better chance of identifying the unique HNL signature, which will be the target of future work.

%%%%%%%%%%%%%%%%%%%% Acknowledgements %%%%%%%%%%%%%%%%%%%%

\begin{acknowledgements}

The IceCube collaboration acknowledges the significant contributions to this manuscript from the Deutsches Elektronen Synchrotron (DESY) and Harvard University groups.
The authors gratefully acknowledge the support from the following agencies and institutions:
USA {\textendash} U.S. National Science Foundation-Office of Polar Programs,
U.S. National Science Foundation-Physics Division,
U.S. National Science Foundation-EPSCoR,
U.S. National Science Foundation-Office of Advanced Cyberinfrastructure,
Wisconsin Alumni Research Foundation,
Center for High Throughput Computing (CHTC) at the University of Wisconsin{\textendash}Madison,
Open Science Grid (OSG),
Partnership to Advance Throughput Computing (PATh),
Advanced Cyberinfrastructure Coordination Ecosystem: Services {\&} Support (ACCESS),
Frontera computing project at the Texas Advanced Computing Center,
U.S. Department of Energy-National Energy Research Scientific Computing Center,
Particle astrophysics research computing center at the University of Maryland,
Institute for Cyber-Enabled Research at Michigan State University,
Astroparticle physics computational facility at Marquette University,
NVIDIA Corporation,
and Google Cloud Platform;
Belgium {\textendash} Funds for Scientific Research (FRS-FNRS and FWO),
FWO Odysseus and Big Science programmes,
and Belgian Federal Science Policy Office (Belspo);
Germany {\textendash} Bundesministerium f{\"u}r Bildung und Forschung (BMBF),
Deutsche Forschungsgemeinschaft (DFG),
Helmholtz Alliance for Astroparticle Physics (HAP),
Initiative and Networking Fund of the Helmholtz Association,
Deutsches Elektronen Synchrotron (DESY),
and High Performance Computing cluster of the RWTH Aachen;
Sweden {\textendash} Swedish Research Council,
Swedish Polar Research Secretariat,
Swedish National Infrastructure for Computing (SNIC),
and Knut and Alice Wallenberg Foundation;
European Union {\textendash} EGI Advanced Computing for research;
Australia {\textendash} Australian Research Council;
Canada {\textendash} Natural Sciences and Engineering Research Council of Canada,
Calcul Qu{\'e}bec, Compute Ontario, Canada Foundation for Innovation, WestGrid, and Digital Research Alliance of Canada;
Denmark {\textendash} Villum Fonden, Carlsberg Foundation, and European Commission;
New Zealand {\textendash} Marsden Fund;
Japan {\textendash} Japan Society for Promotion of Science (JSPS)
and Institute for Global Prominent Research (IGPR) of Chiba University;
Korea {\textendash} National Research Foundation of Korea (NRF);
Switzerland {\textendash} Swiss National Science Foundation (SNSF).

\end{acknowledgements}

\bibliographystyle{apsrev4-2}
\bibliography{hnl_prd_bib}

%%%%%%%%%%%%%%%%%%%% Appendix %%%%%%%%%%%%%%%%%%%%

\clearpage
\appendix

\section{\label{app:signal_shapes}Signal Distributions}

%|U_{\tau4}|^2
% \subsection{Mixing of $\nomathut4=10^{-1}$}
% \subsection{Mixing of $|U_{\tau4}|^2=10^{-1}$}
\subsection{Mixing of \texorpdfstring{$|U_{\tau4}|^2=10^{-1}$}{|U tau4|2=10-1}}

\begin{figure}[h!]
    \includegraphics[width=\columnwidth]{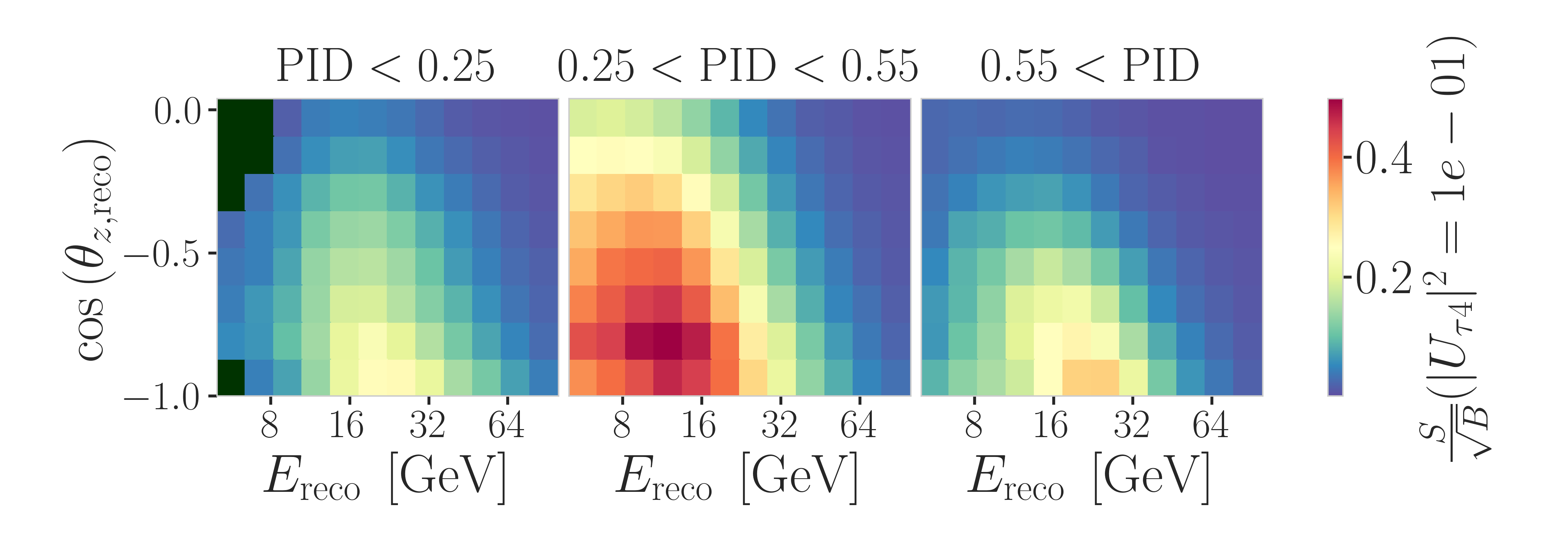}
    \includegraphics[width=\columnwidth]{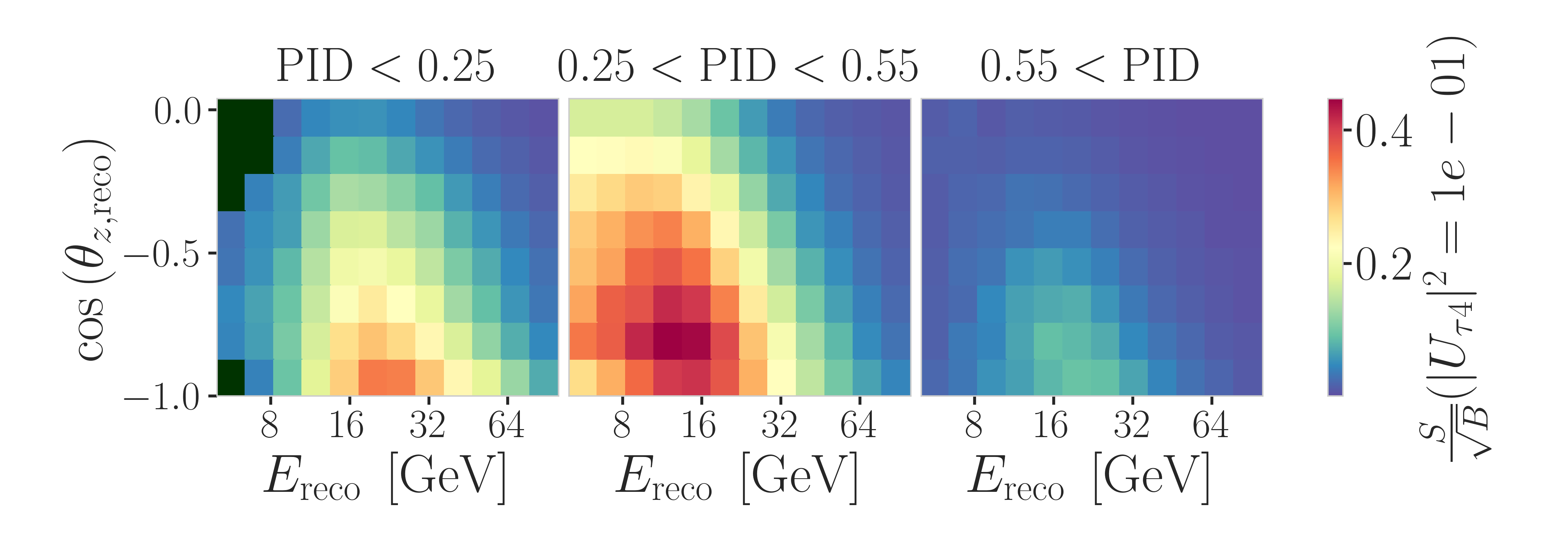}
    \includegraphics[width=\columnwidth]{labeled_s_to_sqrt_b_1_0_GeV_combined_U_tau4_sq_0.1000_total.png}
    \caption{Signal events divided by the square root of the nominal background expectation in \SI{9.28}{years} for all three HNL masses (\SI{0.3}{\gev} top, \SI{0.6}{\gev} middle, \SI{1.0}{\gev} bottom) and a mixing of $\nomathut4=10^{-1}$.}
    \label{fig:s_to_sqrt_b_all_0.1000_mixing}
\end{figure}

\begin{figure}[h!]
    \includegraphics[width=\columnwidth]{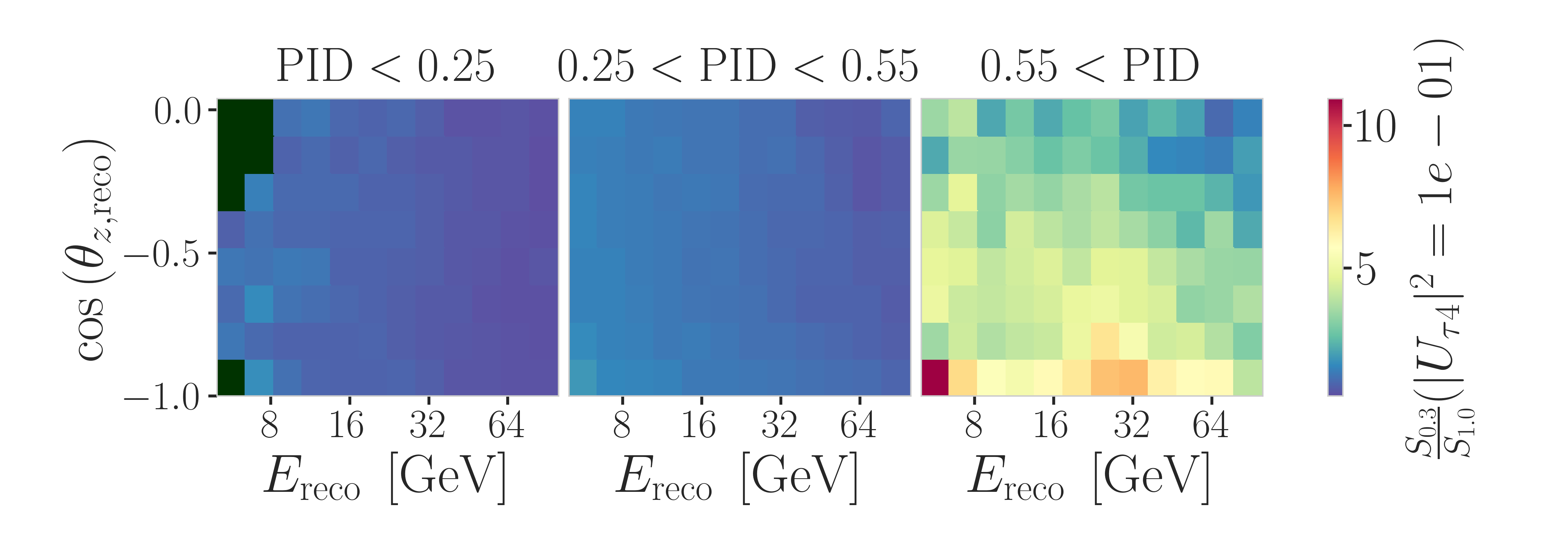}
    \includegraphics[width=\columnwidth]{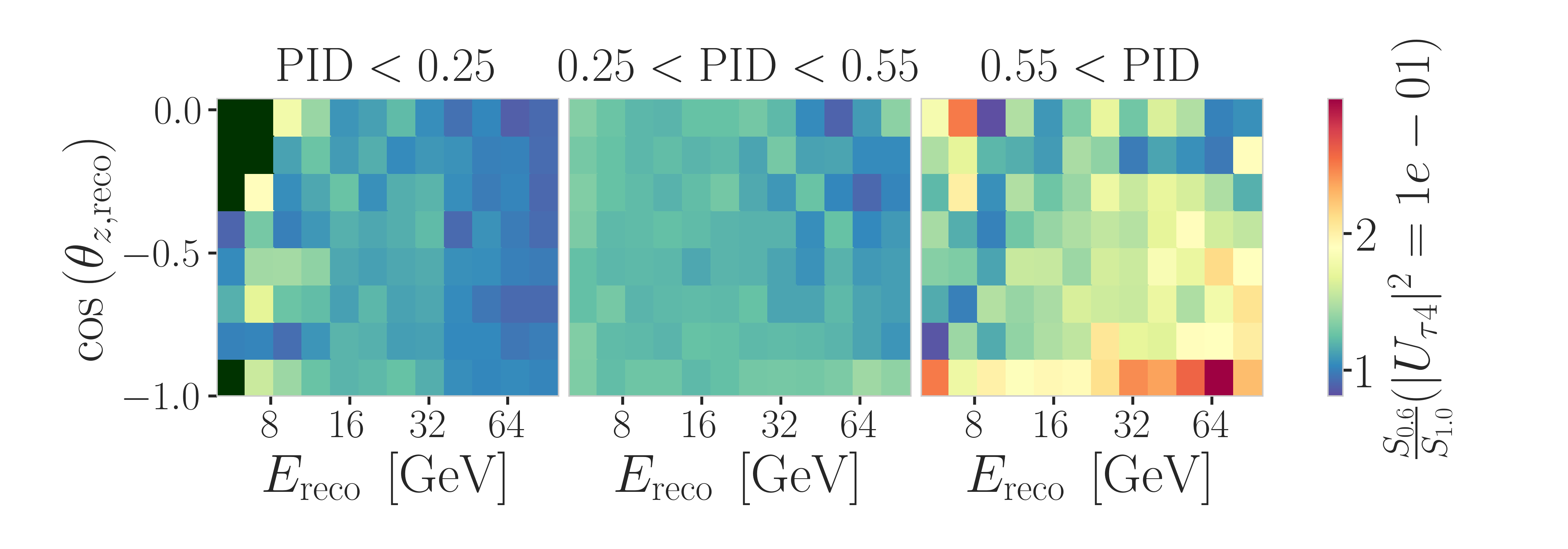}
    \caption{Expected signal events of the \SI{0.3}{\gev} (top) and \SI{0.6}{\gev} (bottom) samples divided by the expected events of the \SI{1.0}{\gev} sample in \SI{9.28}{years} for a mixing of $\nomathut4=10^{-1}$.}
    \label{fig:signal_shape_comparions_0.1000_mixing}
\end{figure}

\newpage

%\subsection{Mixing of $\nomathut4=10^{-3}$}
\subsection{Mixing of \texorpdfstring{$|U_{\tau4}|^2=10^{-3}$}{|U tau4|2=10-3}}

\begin{figure}[h]
    \includegraphics[width=\columnwidth]{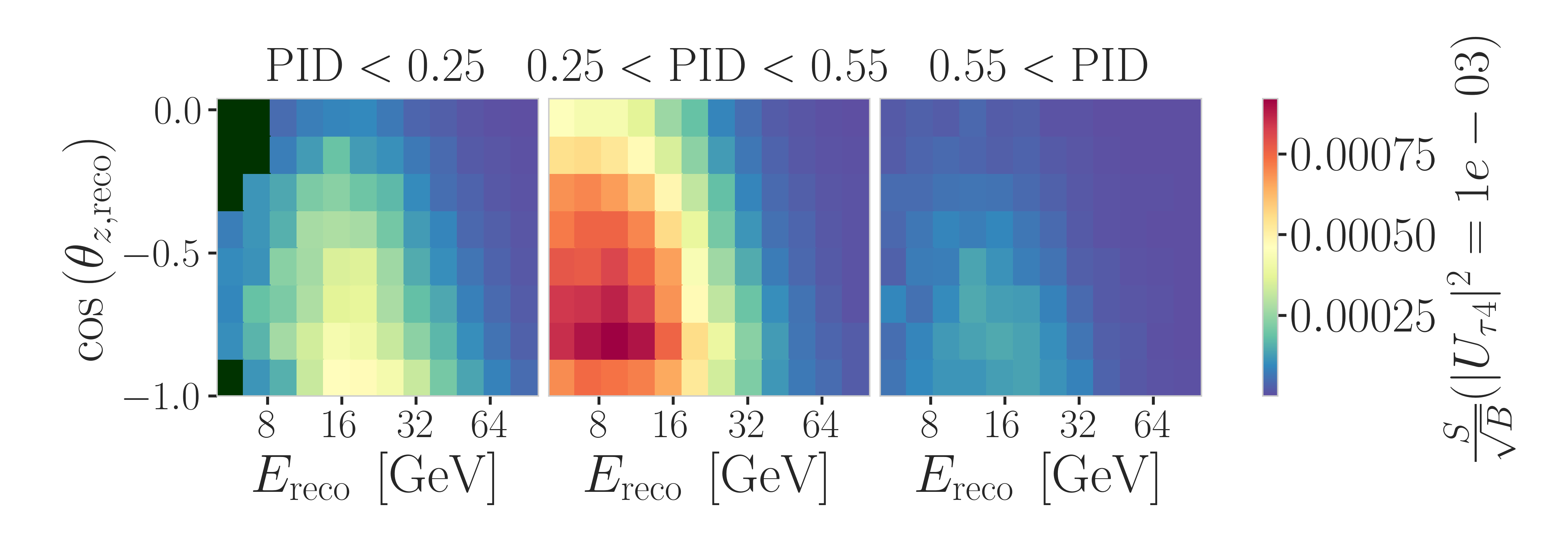}
    \includegraphics[width=\columnwidth]{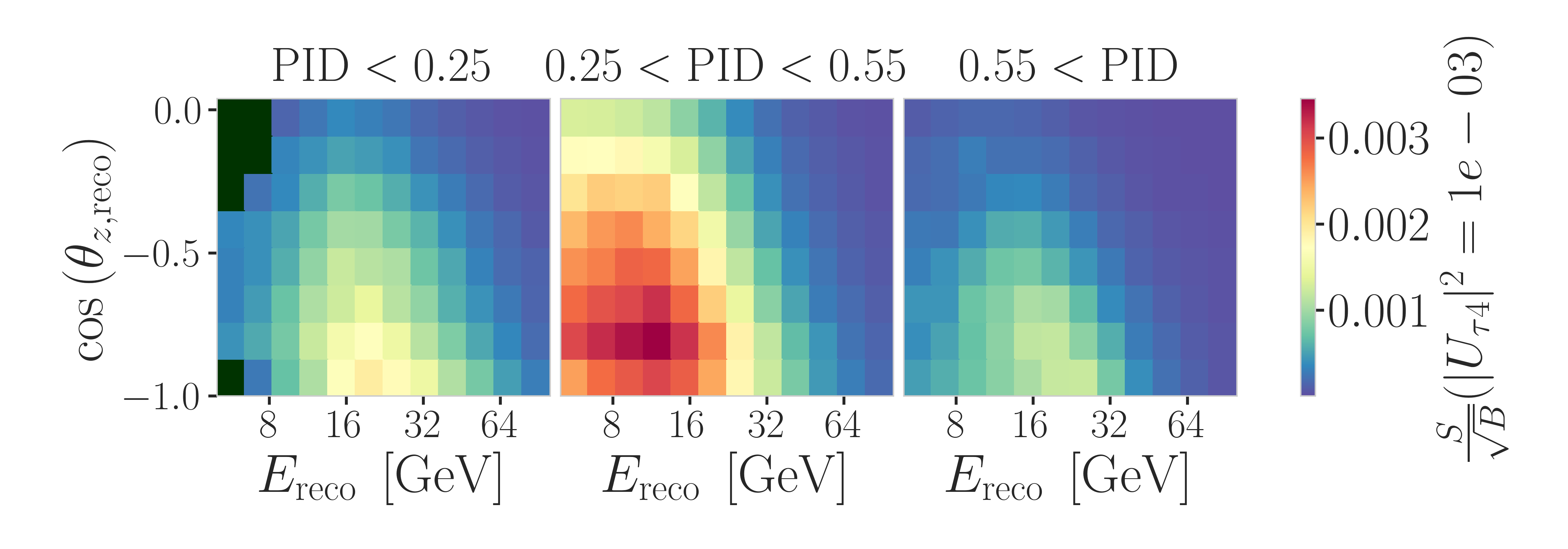}
    \includegraphics[width=\columnwidth]{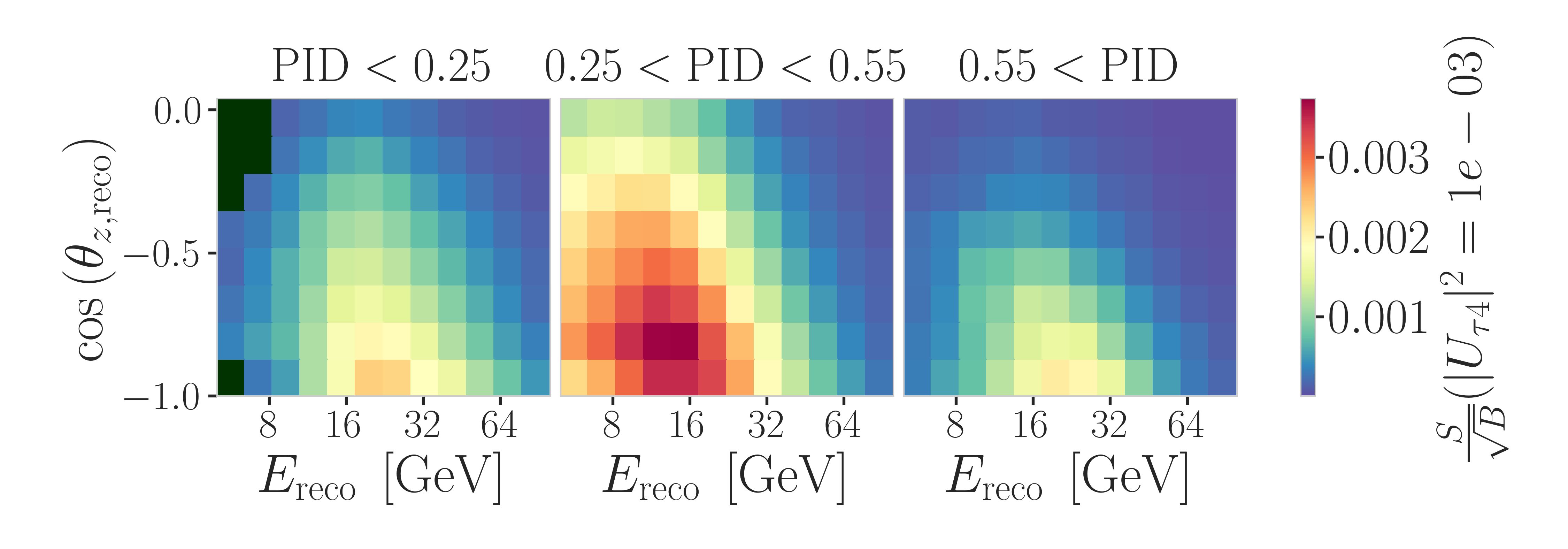}
    \caption{Signal events divided by the square root of the nominal background expectation in \SI{9.28}{years} for all three HNL masses (\SI{0.3}{\gev} top, \SI{0.6}{\gev} middle, \SI{1.0}{\gev} bottom) and a mixing of $\nomathut4=10^{-3}$.}
    \label{fig:s_to_sqrt_b_all_0.0010_mixing}
\end{figure}

\begin{figure}[h]
    \includegraphics[width=\columnwidth]{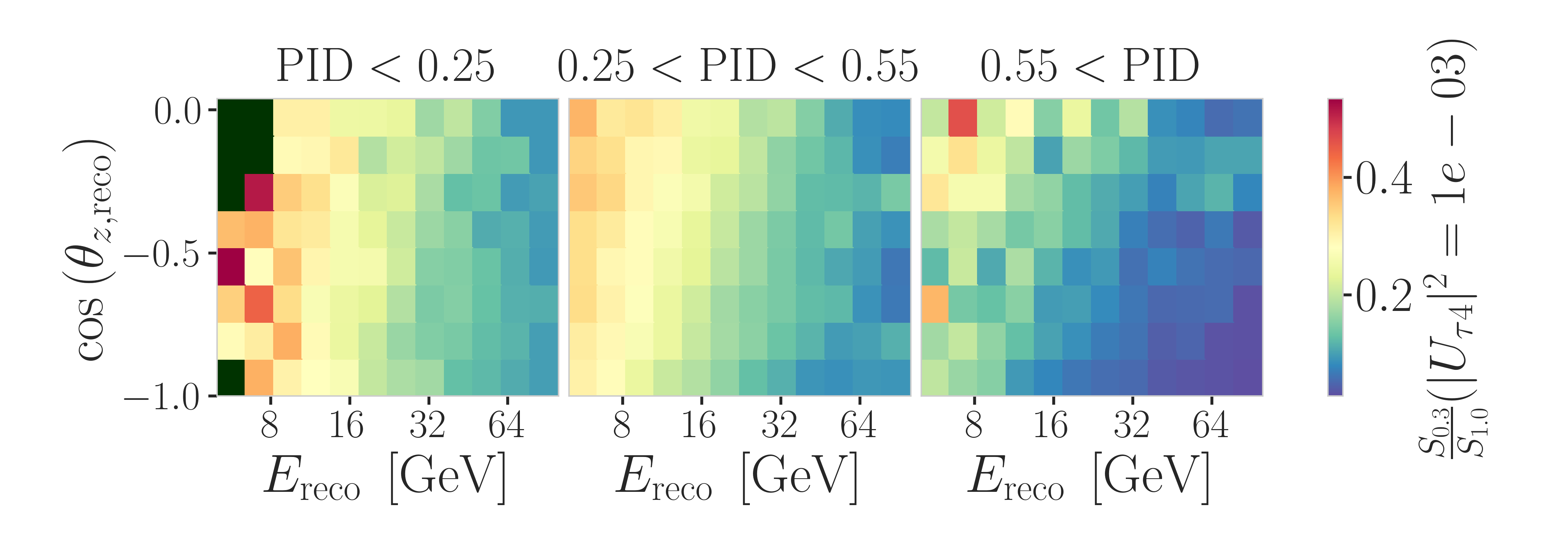}
    \includegraphics[width=\columnwidth]{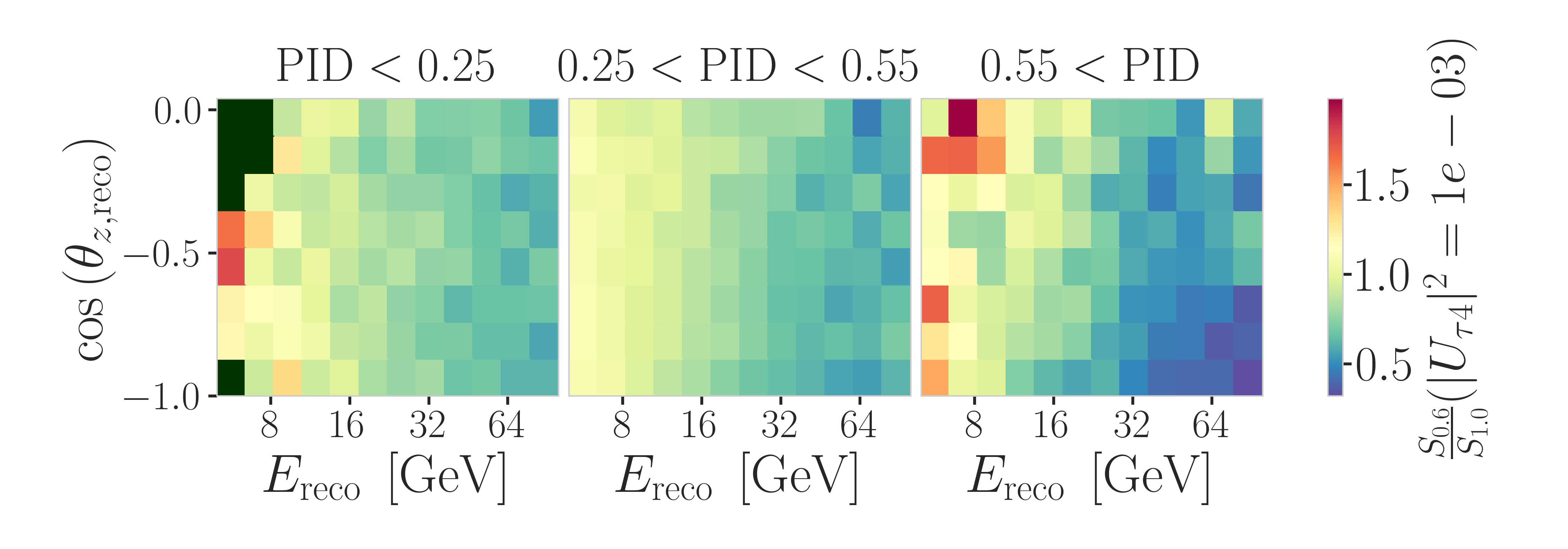}
    \caption{Expected signal events of the \SI{0.3}{\gev} (top) and \SI{0.6}{\gev} (bottom) samples divided by the expected events of the \SI{1.0}{\gev} sample in \SI{9.28}{years} for a mixing of $\nomathut4=10^{-3}$.}
    \label{fig:signal_shape_comparions_0.0010_mixing}
\end{figure}

\end{document}